\documentclass[twocolumn,showpacs,preprintnumbers,amsmath,amssymb]{revtex4}

\usepackage{graphicx} 
\usepackage{dcolumn} 
\usepackage{bm} 

\usepackage{amssymb} 
\usepackage{amsmath} 

\usepackage{here}

\begin{document}


\title{Entropy production due to adiabatic particle creation \\
             in a holographic dissipative cosmology}

\author{Nobuyoshi {\sc Komatsu}}  \altaffiliation{E-mail: komatsu@se.kanazawa-u.ac.jp} 
\affiliation{Department of Mechanical Systems Engineering, Kanazawa University, Kakuma-machi, Kanazawa, Ishikawa 920-1192, Japan}

\date{\today}

\begin{abstract}

Cosmological adiabatic particle creation results in the generation of irreversible entropy.
The evolution of this entropy is examined in a flat Friedmann--Robertson--Walker universe at late times, using a dissipative model with a power-law term (proportional to the power of the Hubble parameter $H$).
In a dissipative universe, the irreversible entropy included in the Hubble volume is found to be proportional to $H^{-1}$, unlike for the case of the Bekenstein--Hawking entropy on the horizon of the universe.
In addition, the evolution of the horizon entropy is examined, extending the previous analysis of a non-dissipative universe [Phys.\ Rev.\ D \textbf{100}, 123545 (2019)].
In the present model, the generalized second law of thermodynamics is always satisfied, whereas the maximization of entropy is satisfied under specific conditions.
The dissipative universe should be constrained by the entropy maximization as if the universe behaves as an ordinary, isolated macroscopic system.
The thermodynamic constraints are likely to be consistent with constraints on a transition from a decelerating universe to an accelerating universe.


\end{abstract}


\maketitle

\section{Introduction} 
\label{Introduction}

An accelerated expansion of the late universe \cite{PERL1998_Riess1998,Planck2018} has been widely accepted as a new paradigm.
To explain the accelerated expansion, astrophysicists have proposed several cosmological models \cite{Bamba1}: 
e.g., $\Lambda$CDM (Lambda cold dark matter) models, $\Lambda (t)$CDM models (i.e., a time-varying $\Lambda(t)$ cosmology) 
\cite{Freese-Mimoso_2015,Sola_2009-2015,Nojiri2006,Sola_2015L14,Valent2015,Sola2019,Sola2020_2},
bulk viscous models \cite{Weinberg0,Barrow11-Zimdahl1,Brevik1-Nojiri1,Barrow21,Avelino2-Brevik}, 
and the creation of CDM (CCDM) models \cite{Prigogine_1988-1989,Lima1992-2016,Lima2014b,Freaza2002,Others2001-2016,Lima2011,Ramos_2014,Ramos_2014b,Koma8,Sola2020,Singha,Cardenas2020}, 
as well as other scenarios \cite{Easson,Cai,Basilakos1,Koma4,Koma5,Basilakos2014-Gohar,Gohar_a_b,Koma6,Koma7,Koma9,Padma2012AB,Cai2012-Tu2013,Tu2013-2015,Neto2018a,Sheykhi1,Sadjadi1,Sheykhi2,Karami2011-2018,Koma1012,Koma11,Koma14}.
The evolution of the universe has been recently examined from a thermodynamic viewpoint, using such models \cite{Sheykhi1,Sadjadi1,Sheykhi2,Karami2011-2018,Koma1012,Koma11,Koma14,Easther1,Davies11_Davis0100,Gong00_01,Egan1,Pavon2013,Mimoso2013,Krishna2017,Krishna2019,Bamba2018,Pavon2019,Saridakis2019,deSitter}.

The formulations of these models can be categorized into two types from a dissipative viewpoint.
The first type is $\Lambda(t)$ \cite{Koma6,Koma14}, which is similar to $\Lambda(t)$CDM models \cite{Freese-Mimoso_2015,Sola_2009-2015,Nojiri2006,Sola_2015L14,Valent2015,Sola2019,Sola2020_2}.
In $\Lambda(t)$ models, both the Friedmann equation and the acceleration equation include an extra driving term \cite{Koma6,Koma14}.
The driving term leads to a non-zero term on the right-hand side of the continuity equation, except for $\Lambda$CDM models.
The non-zero term is considered to be related to `reversible entropy', due to, e.g., the exchange of matter (energy) \cite{Barrow22,Prigogine_1998}.
In this sense, the universe for $\Lambda(t)$ models is non-dissipative.

The second type is BV \cite{Koma6,Koma14}, which is similar to both bulk viscous models \cite{Weinberg0,Barrow11-Zimdahl1,Brevik1-Nojiri1,Barrow21,Avelino2-Brevik} and CCDM models \cite{Prigogine_1988-1989,Lima1992-2016,Lima2014b,Freaza2002,Others2001-2016,Lima2011,Ramos_2014,Ramos_2014b,Koma8,Sola2020,Singha,Cardenas2020}.
In BV (bulk-viscous-cosmology-like) models, the acceleration equation includes an extra driving term, whereas the Friedmann equation does not \cite{Koma6,Koma14}.
This driving term leads to a non-zero term on the right-hand side of the continuity equation even if the driving term is constant (which is similar to $\Lambda$CDM models).
The non-zero term is considered to be related to `irreversible entropy', due to, e.g., gravitationally induced particle creation \cite{Prigogine_1988-1989,Lima1992-2016}.
The universe for BV models is dissipative.

The background evolution of the universe for the $\Lambda(t)$ and BV models is equivalent when the driving terms are the same.
In this case, an associated entropy on the horizon of the universe, e.g., the Bekenstein--Hawking entropy \cite{Bekenstein1Hawking1}, is also equivalent because it depends on the background evolution,
that is, the evolution of the horizon entropy becomes the same in the two models.
However, irreversible entropy due to dissipation is produced in the BV model, whereas it is not produced in the $\Lambda(t)$ model.
An example that has been examined is the irreversible entropy due to adiabatic particle creation; see, e.g., the recent work of Sol\`{a} and Yu \cite{Sola2020}. 

Of course, the irreversible entropy due to adiabatic particle creation should be extremely small compared to the horizon entropy.
However, time derivatives of the entropy play important roles in the second law of thermodynamics and the maximization of entropy \cite{Callen}.
In addition, such a dissipative universe has not yet been systematically examined from a thermodynamic viewpoint, although a non-dissipative universe was examined in a previous work \cite{Koma14}.
Accordingly, it is worth studying the irreversible entropy, in order to clarify the thermodynamic constraints on a dissipative universe.
(The entropy of ordinary, isolated macroscopic systems does not decrease and approaches a certain maximum value in the last stage \cite{Callen}.
A certain type of universe should behave as an ordinary macroscopic system in the last stage, as examined by, e.g., Mimoso and Pav\'{o}n \cite{Mimoso2013}.)

In this context, we study irreversible entropy due to adiabatic particle creation in a flat Friedmann--Robertson--Walker (FRW) universe at late times.
In the present paper, a power-law term is phenomenologically applied to BV models to systematically examine the entropy production in a dissipative universe.
The power-law term \cite{Koma11} can be derived from, e.g., Padmanabhan's holographic equipartition law \cite{Padma2012AB} with a power-law corrected entropy \cite{Das2008Radicella2010}.
Using the dissipative model, we examine the evolution of the irreversible entropy and the Bekenstein--Hawking entropy.

The remainder of the present article is organized as follows.
In Sec.\ \ref{Entropy relation}, an entropy relation for adiabatic particle creation in a flat FRW universe is reviewed.
In Sec.\ \ref{Dissipative model}, a dissipative model that includes a power-law term is formulated.
In Sec.\ \ref{Entropy Sm}, irreversible entropy due to adiabatic particle creation is derived from the entropy relation, using the present model.
In Sec.\ \ref{Entropy evolution}, the evolution of the irreversible entropy and the Bekenstein--Hawking entropy is examined.
The second law of thermodynamics and the maximization of entropy are also discussed.
Finally, in Sec.\ \ref{Conclusions}, the conclusions of the study are presented.

\section{Entropy relation for adiabatic particle creation}
\label{Entropy relation}

Prigogine \textit{et al.} have proposed nonequilibrium thermodynamics of open systems to examine the thermodynamics of cosmological matter creation \cite{Prigogine_1988-1989}. 
Based on this concept, Lima \textit{et al.} studied the radiation temperature law for adiabatic particle creation \cite{Lima1992-2016,Lima2014b}.
The general radiation temperature law in a dissipative universe was investigated by the present author \cite{Koma8}.
Recently, Sol\`{a} and Yu examined entropy production for adiabatic particle creation in a dissipative running-vacuum universe \cite{Sola2020}.
In this section, an entropy relation for adiabatic particle creation is reviewed according to these works.

A spatially flat FRW universe is considered.
The Hubble parameter $H$ is defined by
\begin{equation}
   H \equiv  \frac{ da/dt } {a(t)}  = \frac{ \dot{a}(t) } {a(t)}  ,
\label{eq:Hubble}
\end{equation}
where $a(t)$ is the scale factor at time $t$.
In addition, we consider nonequilibrium thermodynamic states of cosmological fluids in a flat FRW background \cite{Koma8}, assuming adiabatic particle creation \cite{Lima1992-2016,Lima2014b}.
The balance equations for the number of particles, entropy, and energy can be written as
\begin{equation}
       \dot{n} + 3 H n   = n \Gamma   ,    
\label{eq:NonEquil_1}
\end{equation}
\begin{equation}
       \dot{s} + 3 H s   = s \Gamma   ,    
\label{eq:NonEquil_2}
\end{equation}
\begin{equation}
       \dot{\varepsilon} + 3 H ( \varepsilon + p + p_{c} )   = 0   ,
\label{eq:NonEquil_3}
\end{equation}
where $n$, $s$, $\varepsilon$, and $p$ are the particle number density, entropy density, energy density, and pressure, respectively \cite{Lima2014b}. 
$\Gamma$ and $p_{c}$ are the particle production rate and the dynamic creation pressure, respectively.
The three balance equations reduce to the conservation law for equilibrium states in a standard cosmology when both $\Gamma =0$ and $p_{c}=0$ \cite{Lima2014b,Koma8}.

The total number $N$ of particles and the entropy $S$ in the comoving volume can be given by \cite{Lima2014b}
\begin{equation}
        N = n a^3   \quad \textrm{and} \quad  S = s a^3   .
\label{eq:NS-a3}
\end{equation}
Accordingly, Eq.\ (\ref{eq:NonEquil_1}) is written as 
\begin{equation}
\frac{\dot{N}}{N}  = \Gamma   .
\label{eq:N-Gamma}
\end{equation}
In this paper, the entropy per particle $\sigma \equiv S/N$ is assumed to be constant \cite{Lima1992-2016,Lima2014b,Sola2020}:
\begin{equation}
    \sigma \equiv \frac{S}{N} = \textrm{cst.} \quad \textrm{or equivalently} \quad   \dot{\sigma} =0. 
\label{eq:sigma_cst}
\end{equation}
The constant $\sigma$ has been used for Eq.\ (\ref{eq:NonEquil_2}).
From Eqs.\ (\ref{eq:NS-a3}), (\ref{eq:N-Gamma}), and (\ref{eq:sigma_cst}), Eq.\ (\ref{eq:NonEquil_2}) is rewritten as 
\begin{equation}
\frac{\dot{S}}{S}  = \frac{\dot{N}}{N} + \frac{\dot{\sigma}}{\sigma}  = \Gamma+ \frac{\dot{\sigma}}{\sigma} =\Gamma   , 
\label{eq:S-Gamma}
\end{equation}
where $N \neq 0$ and $S \neq 0$ are assumed \cite{Sola2020}.

An entropy relation for adiabatic particle creation is calculated from Eqs.\ (\ref{eq:NonEquil_2}) and (\ref{eq:S-Gamma}).
For example, reformulating Eq.\ (\ref{eq:NonEquil_2}), we obtain 
\begin{equation}
\frac{\dot{s}}{s}  = \Gamma -3H  .
\label{eq:s-Gamma}
\end{equation}
Integrating Eq.\ (\ref{eq:s-Gamma}) from the present time $t_{0}$ to an arbitrary time $t$ gives
\begin{equation}
     \int_{s_{0}}^{s}  \frac{d s^{\prime} }{ s^{\prime} }  =  \int_{t_{0}}^{t}  ( \Gamma (t^{\prime})  - 3H(t^{\prime})   )  dt^{\prime}  ,
\label{eq:dots_s_int1}
\end{equation}
and solving this equation yields
\begin{equation}
     \frac{s}{s_{0}}  = \exp  \left [  \int_{t_{0}}^{t}   ( \Gamma (t^{\prime})  - 3H(t^{\prime})   )  dt^{\prime} \right ]  ,
\label{eq:dots_s_s-t}
\end{equation}
where $s_{0}$ is the entropy density at the present time.
Transforming an integral parameter from $t^{\prime}$ to $\tilde{a}^{\prime}$, and using $H = \dot{a}/a = \dot{\tilde{a}} /\tilde{a}$, Eq.\ (\ref{eq:dots_s_s-t}) can be written as 
\begin{align}
    \frac{s}{s_{0}}  &= \exp  \left [  \int_{1}^{{\tilde{a}}}   (  \Gamma (\tilde{a}^{\prime}) -3H(\tilde{a}^{\prime})  )  \frac{ dt^{\prime} }{ d\tilde{a}^{\prime} } d\tilde{a}^{\prime} \right ]                 \notag \\
                         &= \exp  \left [  \int_{1}^{{\tilde{a}}}   (  \Gamma (\tilde{a}^{\prime}) -3H(\tilde{a}^{\prime})  )  \frac{ \tilde{a}^{\prime} }{ \dot{\tilde{a}}^{\prime} } \frac{ d\tilde{a}^{\prime} }{\tilde{a}^{\prime} } \right ]  \notag \\
                         &= \exp  \left [  \int_{1}^{{\tilde{a}}}   \left ( \frac{ \Gamma (\tilde{a}^{\prime}) }{ H }  -3 \right )  \frac{ d\tilde{a}^{\prime} }{\tilde{a}^{\prime} } \right ]  ,
\label{eq:dots_s_s-a}
\end{align}
where $\tilde{a}$ is the normalized scale factor given by 
\begin{equation}
   \tilde{a} = \frac{a} { a_{0}}      ,
\label{eq:a_a0}
\end{equation}
and $a_{0}$ is the scale factor at the present time.
Equation\ (\ref{eq:dots_s_s-a}) is an entropy density relation for adiabatic particle creation.
The evolution of the entropy density depends on the particle production rate and the background evolution of the universe.
In the next section, we discuss the background evolution of the universe in a dissipative model that includes a power-law term.

Before proceeding further, we discuss the balance equation for the energy density, which is shown in Eq.\ (\ref{eq:NonEquil_3}).
The local Gibbs relation should be valid in the nonequilibrium thermodynamic states considered here \cite{Lima2014b}.
The local Gibbs relation can be written as
\begin{equation}
         n k_{B} T d \left ( \frac{s}{n}    \right ) \equiv  n k_{B} T d\sigma  = d \varepsilon - \frac{\varepsilon + p}{n} dn      ,
\label{eq:localGibbs}
\end{equation}
where $k_{B}$ and $T$ are the Boltzmann constant and the temperature, respectively.
Substituting $\dot{\sigma} =d\sigma/dt =0$ into Eq.\ (\ref{eq:localGibbs}) and applying the resultant equation and Eq.\ (\ref{eq:NonEquil_1}) to Eq.\ (\ref{eq:NonEquil_3}), we obtain the dynamic creation pressure given by \cite{Koma8}
\begin{equation}
        p_{c}  = - (\varepsilon + p) \frac{\Gamma}{3 H}  .    
\label{eq:pc}
\end{equation}
From this relation, the balance equation for energy given by Eq.\ (\ref{eq:NonEquil_3}) is rewritten as
\begin{equation}
       \dot{\varepsilon} + 3 H ( \varepsilon + p  )   = (\varepsilon + p) \Gamma   .
\label{eq:NonEquil_3_sigma-cst}
\end{equation}
In a matter-dominated universe, i.e., $p=0$, the above equation is 
\begin{equation}
       \dot{\varepsilon} + 3 H  \varepsilon   = \varepsilon  \Gamma   .
\label{eq:NonEquil_3_sigma-cst_p=0}
\end{equation}
Using the mass density $\rho =\varepsilon /c^{2}$, we have
\begin{equation}
       \dot{\rho} + 3 H  \rho   = \rho  \Gamma   ,
\label{eq:NonEquil_3_sigma-cst_p=0_mass}
\end{equation}
where $c$ is the speed of light.
Equation\ (\ref{eq:NonEquil_3_sigma-cst_p=0_mass}) is used in the next section.

\section{Dissipative cosmological model in a flat FRW universe} 
\label{Dissipative model}

In this section, a dissipative model that includes a power-law term is formulated, to systematically examine irreversible entropy in a dissipative universe.
In Sec.\ \ref{General cosmological equations}, we review cosmological equations in a flat FRW universe for $\Lambda(t)$ and BV models.
In Sec.\ \ref{BV model with a power-law term}, we formulate the BV model with a power-law term.
We assume an expanding universe from observations \cite{Hubble2017}.

\subsection{Cosmological equations for $\Lambda(t)$ and BV models} 
\label{General cosmological equations}

We review cosmological equations for $\Lambda(t)$ and BV models, according to Refs.\ \cite{Koma9,Koma14}.
The Friedmann, acceleration, and continuity equations are written as 
\begin{equation}
 H(t)^2      =  \frac{ 8\pi G }{ 3 } \rho (t)    + f_{\Lambda}(t)            ,                                                 
\label{eq:General_FRW01_f_0} 
\end{equation} 
\begin{align}
  \frac{ \ddot{a}(t) }{ a(t) }   
                                          &=  -  \frac{ 4\pi G }{ 3 }  ( 1+  3w ) \rho (t)                                   +   f_{\Lambda}(t)    +  h_{\textrm{B}}(t)  , 
\label{eq:General_FRW02_g_0}
\end{align}
\begin{equation}
       \dot{\rho} + 3  H (1+w)  \rho   =    -  \frac{3\dot{f}_{\Lambda}(t)}{8 \pi G }       +    \frac{3 H h_{\textrm{B}}(t) }{4 \pi G}                  , 
\label{eq:drho_General_00_ri}
\end{equation}
where $G$ is the gravitational constant \cite{Koma9}.
$w$ represents the equation of state parameter for a generic component of matter, $ w =  p / (\rho  c^2) $.
Two extra driving terms, $f_{\Lambda}(t)$ and $h_{\textrm{B}}(t)$, are phenomenologically assumed \cite{Koma14}.

In the above formulation, $f_{\Lambda}(t)$ is used for $\Lambda (t)$ models and $h_{\textrm{B}}(t)$ is used for BV models \cite{Koma9,Koma14}.
Accordingly, we set $h_{\textrm{B}}(t) =0$ for the $\Lambda (t)$ model and $f_{\Lambda}(t) = 0$ for the BV model.
In addition, $f_{\Lambda}(t)$ and $h_{\textrm{B}}(t)$ are assumed to be related to reversible and irreversible processes, respectively.
That is, the BV model assumes an irreversible entropy arising from dissipative processes such as particle creation \cite{Prigogine_1988-1989,Lima1992-2016}.
Based on this assumption, the dynamic creation pressure $p_{c}$ given by Eq.\ (\ref{eq:pc}) is considered to be related to the irreversible entropy.
In contrast, the $\Lambda (t)$ model assumes a reversible entropy, such as that related to the reversible exchange of matter (energy) \cite{Prigogine_1998}. 
Consequently, the first term on the right-hand side of Eq.\ (\ref{eq:drho_General_00_ri}) is related to the reversible entropy
and the second term is related to the irreversible entropy.
We note that it can of course be assumed that $f_{\Lambda}(t)$ and $h_{\textrm{B}}(t)$ are based on other mechanisms, rather than the reversible and irreversible processes assumed in this paper.

In the present study, a matter-dominated universe, i.e., $w =0$, is considered.
Coupling Eq.\ (\ref{eq:General_FRW01_f_0}) with Eq.\ (\ref{eq:General_FRW02_g_0}) yields \cite{Koma14} 
\begin{equation}
    \dot{H} = - \frac{3}{2}  H^{2}  +  \frac{3}{2}    f_{\Lambda}(t)     + h_{\textrm{B}}(t)   .  
\label{eq:Back2}
\end{equation}
Using this equation, we examine the background evolution of the universe in various cosmological models.
Hereafter, we consider the BV model to examine a dissipative universe.

It should be noted that the background evolution of the universe in the $\Lambda(t)$ and BV models is equivalent if the driving terms are equal, i.e., $\frac{3}{2}  f_{\Lambda}(t) = h_{\textrm{B}}(t)$.
However, even in this case, density perturbations related to structure formation are different because the right-hand side of the continuity equation is different, as shown in Eq.\ (\ref{eq:drho_General_00_ri}).
For example, a constant $h_{\textrm{B}}(t)$ leads to a non-zero term on the right-hand side, whereas a constant $f_{\Lambda}(t)$ does not.
For $\Lambda(t)$ models, see, e.g., the works of Sol\`{a} \textit{et al.} \cite{Sola_2015L14}, G\'{o}mez-Valent \textit{et al.} \cite{Valent2015}, and Rezaei \textit{et al.} \cite{Sola2019}.
For BV models, see, e.g., the works of Li and Barrow \cite{Barrow21}, Jesus \textit{et al.} \cite{Lima2011}, and Ramos \textit{et al.} \cite{Ramos_2014,Ramos_2014b}.
In this study, density perturbations are not discussed.

\subsection{BV model with a power-law term} 
\label{BV model with a power-law term}

We phenomenologically formulate a dissipative model that includes a power-law term that is proportional to the power of $H$.
The power-law term is briefly reviewed here.

Based on Padmanabhan's holographic equipartition law with an associated entropy on the horizon, cosmological equations can be derived from the expansion of cosmic space due to the difference between the degrees of freedom on the surface and in the bulk \cite{Padma2012AB}. 
The emergence of the cosmological equation has been examined from various viewpoints \cite{Cai2012-Tu2013,Tu2013-2015,Neto2018a,Koma1012,Koma11}.
In particular, in Ref.\ \cite{Koma11}, an acceleration equation that includes $H^{\alpha}$ terms is derived using the holographic equipartition law \cite{Padma2012AB} with a power-law corrected entropy \cite{Das2008Radicella2010}.
Here, $\alpha$ is a real number and is assumed to be related to the entanglement of quantum fields inside and outside the horizon \cite{Koma11,Das2008Radicella2010}. 
The power-law term has been investigated in a non-dissipative universe based on $\Lambda (t)$ models \cite{Koma11,Koma14}.
The derived acceleration equation may imply that the $H^{\alpha}$ term can be applied to $h_{\textrm{B}}(t)$ for BV models.
In addition, a similar $H^{\alpha}$ term has been examined in CCDM models, see, e.g., the works of Freaza \textit{et al.} \cite{Freaza2002}, Ramos \textit{et al.} \cite{Ramos_2014}, and C\'{a}rdenas \textit{et al.} \cite{Cardenas2020}.

In this context, the power-law term, i.e., the $H^{\alpha}$ term, is applied to BV models, in order to systematically examine a dissipative universe.
Accordingly, the two driving terms $f_{\Lambda}(t)$ and $h_{\textrm{B}}(t)$ are set to be \cite{Koma11,Koma14}
\begin{equation}
        f_{\Lambda}(t)   =   0,
\label{eq:fL}
\end{equation}
\begin{equation}
 h_{\textrm{B}}(t) =   \Psi_{\textrm{B}} H_{0}^{2} \left (  \frac{H}{H_{0}} \right )^{\alpha}  .
\label{eq:hB}
\end{equation}
Here, $H_{0}$ represents the Hubble parameter at the present time.
$\alpha$ and $\Psi_{\textrm{B}}$ are dimensionless constants whose values are real numbers.
In the present paper, $\alpha$ and $\Psi_{\textrm{B}}$ are considered to be independent free parameters \cite{Koma14}.
That is, we phenomenologically assume the power-law term, without using a covariant action.
(In this sense, the BV model considered here should be effective models based on macroscopic thermodynamic properties. 
For microscopic models, see, e.g., $\Lambda(t)$ models based on quantum field theory and string theory \cite{Sola2020_2}.)

Substituting Eqs.\ (\ref{eq:fL}) and (\ref{eq:hB}) into Eq.\ (\ref{eq:Back2}) yields
\begin{align}
    \dot{H} &= - \frac{3}{2} H^{2}  +  \Psi_{\textrm{B}}  H_{0}^{2} \left (  \frac{H}{H_{0}} \right )^{\alpha}      \notag \\
               &= - \frac{3}{2} H^{2}  \left (  1 -   \frac{2}{3} \Psi_{\textrm{B}} \left (  \frac{H}{H_{0}} \right )^{\alpha -2} \right )         \notag \\
               &= - \frac{3}{2} H^{2}  \left (  1 -   \Psi_{\alpha} \left (  \frac{H}{H_{0}} \right )^{\alpha -2} \right )      , 
\label{eq:Back_power_hB}
\end{align}
where $\Psi_{\textrm{B}}$ is replaced by $\Psi_{\alpha}$, a density parameter for the effective dark energy \cite{Koma14}, which is written as
\begin{equation}
 \Psi_{\alpha} =   \frac{2}{3} \Psi_{\textrm{B}}   .
\label{eq:Psi_a-B}
\end{equation}
In addition, the following is assumed for $\Psi_{\alpha}$, 
\begin{equation}
       0 \leq \Psi_{\alpha} \leq 1 .
\label{eq:Psi_01}
\end{equation}
The formulation of Eq.\ (\ref{eq:Back_power_hB}) is equivalent to that examined in a previous work \cite{Koma14}.
Accordingly, using the result in Ref.\ \cite{Koma14}, the solution is written as 
\begin{equation}  
    \left ( \frac{H}{H_{0}} \right )^{2-\alpha}  =   (1- \Psi_{\alpha})   \tilde{a}^{ - \frac{3(2-\alpha)}{2}  }  + \Psi_{\alpha}   ,
\label{eq:Sol_HH0_power}
\end{equation}
where $\tilde{a}$ is a normalized scale factor given by Eq.\ (\ref{eq:a_a0}).
Note that equations for $\alpha \neq 2$ are shown in this paper because when $\alpha \rightarrow 2$, they reduce to those for $\alpha = 2$.
For example, Eq.\ (\ref{eq:Sol_HH0_power}) reduces to $\frac{H}{H_{0}}  = \tilde{a}^{ - \frac{3 (1- \Psi_{\alpha}) }{2}  }$ when $\alpha \rightarrow 2$ \cite{Koma14}.

The background evolution of the universe in the present dissipative model is calculated from Eq.\ (\ref{eq:Sol_HH0_power}).
In particular, when $\alpha =0$, replacing $\Psi_{\alpha}$ by $\Omega_{\Lambda}$, the density parameter for $\Lambda$, gives a background evolution that is equivalent to that of a non-dissipative universe in $\Lambda$CDM models. 
(The density parameter for matter is given by $1- \Omega_{\Lambda}$, neglecting the influence of radiation \cite{Koma14}, in a flat FRW universe at late times.)

The temporal deceleration parameter $q$ is also useful for examining the background evolution of the universe. 
The deceleration parameter is defined by  
\begin{equation}
q \equiv  - \left ( \frac{\ddot{a} } {a H^{2}} \right )  , 
\label{eq:q_def}
\end{equation}
where a positive and negative $q$ represent deceleration and acceleration, respectively \cite{Koma14}. 
Substituting $\ddot{a}/a = \dot{H} + H^{2}$ into Eq.\ (\ref{eq:q_def}) and applying Eqs.\ (\ref{eq:Back_power_hB})  and (\ref{eq:Sol_HH0_power}) to the resultant equation yields
\begin{align}
q  &=   -  \frac{ \dot{H} } {H^{2}}   -1     =\frac{3}{2}    \left (  1 -   \Psi_{\alpha} \left (  \frac{H}{H_{0}} \right )^{\alpha -2} \right )             -1     \notag \\
    &= \frac{1}{2}  -   \frac{3}{2}     \Psi_{\alpha} \left [  (1- \Psi_{\alpha})   \tilde{a}^{ - \frac{3(2-\alpha)}{2}  }  + \Psi_{\alpha}    \right ]^{-1}    .
\label{eq:q_power}
\end{align}
In addition, from this equation, the boundary for $q = 0$ can be written as \cite{Koma14}
\begin{align}  
   \Psi_{\alpha} =  \frac{  \tilde{a}^{ - \frac{3(2-\alpha)}{2}  }       }{  2  +  \tilde{a}^{ - \frac{3(2-\alpha)}{2}  }       }  ,
\label{eq:q=0}
\end{align}
or equivalently, 
\begin{align}  
  \tilde{a} = \left (  \frac{2 \Psi_{\alpha} } { 1 - \Psi_{\alpha} }   \right )^{ - \frac{2}{3(2-\alpha)}   }       .
\label{eq:q=0_a}
\end{align}
Using the boundary in the $(\Psi_{\alpha}, \alpha)$ plane, we can discuss a transition between deceleration and acceleration.
We discuss this in Sec.\ \ref{Entropy evolution}.

\begin{figure} [t] 
\begin{minipage}{0.495\textwidth}
\begin{center}
\scalebox{0.33}{\includegraphics{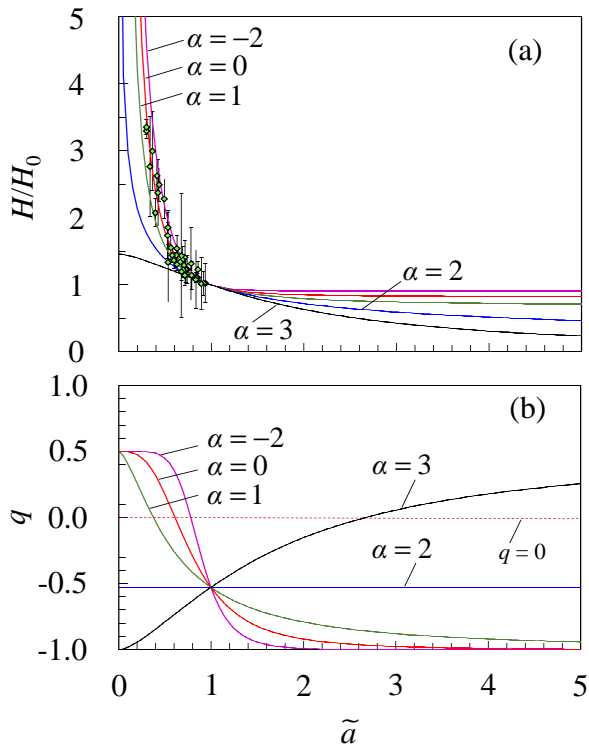}}
\end{center}
\end{minipage}
\caption{ (Color online). Evolution of the universe for the dissipative model for $\Psi_{\alpha} =0.685$.
(a) Normalized Hubble parameter $H/H_{0}$.
(b) Deceleration parameter $q$.
In (a), the closed diamonds with error bars are observed data points taken from Ref.\ \cite{Hubble2017}. 
To normalize the data points, $H_{0}$ is set to $67.4$ km/s/Mpc from Ref.\ \cite{Planck2018}. 
In (b), the horizontal break line represents $q=0$.
A similar evolution has been examined for a non-dissipative universe \cite{Koma14}.
The background evolution of the dissipative universe is essentially equivalent to that of the non-dissipative universe. See the text. }
\label{Fig-H-a}
\end{figure}

To observe the background evolution of the dissipative universe, the evolutions of the Hubble parameter and the deceleration parameter are shown in Fig.\ \ref{Fig-H-a}.
To examine typical results, $\alpha$ is set to $-2$, $0$, $1$, $2$, and $3$.
In addition, $\Psi_{\alpha}$ is set to $0.685$, which is equivalent to $\Omega_{\Lambda}$ for the $\Lambda$CDM model from the Planck 2018 results \cite{Planck2018}.
That is, the background evolution of the dissipative universe examined here is set to be the same as that of a non-dissipative universe examined in Ref.\ \cite{Koma14}.
 (A similar evolution of a non-dissipative universe has been previously discussed \cite{Koma14}.) 
In this study, the normalized scale factor $\tilde{a}$ increases with time because an expanding universe is considered.

As shown in Fig.\ \ref{Fig-H-a}(a), for all $\alpha$, $H/H_{0}$ is $1$ at the present time and decreases with $\tilde{a}$.
For $\alpha <2$, $H/H_{0}$ gradually approaches a positive value, whereas for $\alpha = 3$, it gradually approaches $0$ \cite{Koma14}.
Similarly, for all $\alpha$, $q$ is negative at the present time [Fig.\ \ref{Fig-H-a}(b)].
The negative value is $-0.5275$, which is calculated from Eq.\ (\ref{eq:q_power}).
A negative $q$ represents an accelerating universe.
In addition, $q$ for $\alpha <2$ decreases with $\tilde{a}$ and gradually approaches $-1$, although it is positive in the early stage.
This result indicates an initially decelerating and then accelerating universe (hereafter `decelerating and accelerating universe').
From Eq.\ (\ref{eq:q=0_a}), the transition points for $\alpha = -2$, $0$, and $1$ are approximately $\tilde{a} =0.783$ ($z=0.278$), $\tilde{a}=0.613$ ($z=0.632$), and $\tilde{a}=0.375$ ($z=1.664$), respectively.
Here $z$ represents the redshift, which is given by $z= \tilde{a}^{-1} -1$.
Note that the transition point for $\alpha =0$, i.e., $z=0.632$, corresponds to that for the $\Lambda$CDM model.
In contrast, the evolution of $q$ for $\alpha= 3$ indicates an initially accelerating and then decelerating universe (hereafter `accelerating and decelerating universe') [Fig.\ \ref{Fig-H-a}(b)].
Therefore, we can expect that $\alpha <2$ corresponds to the decelerating and accelerating universe, whereas $\alpha >2$ corresponds to the accelerating and decelerating universe.
This expectation is examined in Sec.\ \ref{Entropy evolution}.

The above result is consistent with that for a non-dissipative universe examined in Ref.\ \cite{Koma14}, because the background evolution is set to be the same as that for the non-dissipative universe.
However, irreversible entropy due to adiabatic particle creation is produced in the dissipative universe, unlike in the non-dissipative universe.
To examine the irreversible entropy, we calculate the relationship between the driving term $h_{\textrm{B}}(t)$ and the particle production rate $\Gamma$.
Substituting $ f_{\Lambda}(t) =0$ and $w=0$ into Eq.\ (\ref{eq:drho_General_00_ri}), the continuity equation is given by 
\begin{equation}
       \dot{\rho} + 3  H   \rho   =    \frac{3 H h_{\textrm{B}}(t) }{4 \pi G}                  .
\label{eq:drho_General_00_ri_w=0}
\end{equation}
From Eqs.\ (\ref{eq:NonEquil_3_sigma-cst_p=0_mass}) and (\ref{eq:drho_General_00_ri_w=0}), $\Gamma$ is written as
\begin{equation}
       \Gamma  =    \frac{3 H }{4 \pi G}    \frac{ h_{\textrm{B}}(t) }{ \rho }                , 
\label{eq:Gamma_w=0_hb}
\end{equation}
where $p=0$ and $\dot{\sigma}=0$ have been assumed.
This equation indicates that $\Gamma$ depends on $H$, $\rho$, and $h_{\textrm{B}}(t)$.
In the next section, we examine the irreversible entropy in the dissipative model, using this equation.

It should be noted that $\alpha$ greatly affects a transition point of $q=0$.
As shown in Fig.\ \ref{Fig-H-a}(b), the evolutions of $q$ for $\alpha =-2$, $0$, and $1$ satisfy a decelerating and accelerating universe.
However, the transition points for $\alpha = -2$ and $1$ significantly deviate from that point for $\alpha = 0$, corresponding to the $\Lambda$CDM model from the Planck 2018 results \cite{Planck2018}.
This deviation implies that $\alpha = -2$ and $1$ are not consistent with an observed transition point.
Accordingly, the range of $\alpha$ is expected to be further constrained by observations, using the transition point.
In fact, the observations imply that small $|\alpha|$, such as $|\alpha| < 1$, is favored.
(The transition point for $\alpha = 0$ is $z=0.632$, as examined above.)
For example, the transition point for $\alpha = 0.38$, i.e., $z=0.831$, should be an upper limit which satisfies $z=0.632 \pm 0.200$, if observation errors are assumed to be $\Delta z = \pm 0.200$.
In this way, constraints on $\alpha$ can be estimated from the observed data.
The estimated range of $\alpha$ depends on the accuracy of the observation.
The exact observational constraints on $\alpha$ are not discussed in the present study.

\section{Irreversible entropy $S_{m}$ for the present dissipative model} 
\label{Entropy Sm}

In this section, we examine an irreversible entropy $S_{m}$ due to adiabatic particle (matter) creation for the present dissipative model.
In Sec.\ \ref{Entropy density}, the entropy density is calculated from an entropy density relation.
In Sec.\ \ref{Entropy in the Hubble volume}, the entropy $S_{m}$ in the Hubble volume is derived from the entropy density.
Note that the Hubble volume is the volume of the sphere with the Hubble horizon, as described later.

\subsection{Entropy density} 
\label{Entropy density}

The entropy density is calculated from an entropy density relation examined in Sec. \ref{Entropy relation}.
From Eq.\ (\ref{eq:dots_s_s-a}), the entropy density relation is written as
\begin{align}
    \frac{s}{s_{0}}    &= \exp  \left [  \int_{1}^{{\tilde{a}}}   \left ( \frac{ \Gamma (\tilde{a}^{\prime}) }{ H }  -3 \right )  \frac{ d\tilde{a}^{\prime} }{\tilde{a}^{\prime} } \right ]  .
\label{eq:dots_s_s-a_2}
\end{align}

We now calculate Eq.\ (\ref{eq:dots_s_s-a_2}) using the present dissipative model.
For this, we first calculate $\Gamma /H$.
From Eq.\ (\ref{eq:Gamma_w=0_hb}), $\Gamma /H$ is written as
\begin{equation}
     \frac{  \Gamma }{ H }  =    \frac{3}{4 \pi G}    \frac{ h_{\textrm{B}}(t) }{ \rho }                .
\label{eq:Gamma-H_w=0_hb}
\end{equation}
Substituting $ h_{\textrm{B}}(t) $ given by Eq.\ (\ref{eq:hB}) and $\rho$ given by Eq.\ (\ref{eq:General_FRW01_f_0}) into Eq.\ (\ref{eq:Gamma-H_w=0_hb}) yields 
\begin{align}
     \frac{  \Gamma }{ H }  =    \frac{3}{4 \pi G}  \frac{  \Psi_{\textrm{B}} H_{0}^{2} \left (  \frac{H}{H_{0}} \right )^{\alpha}   }{ \frac{3 H^{2} }{ 8\pi G }  }    
                                      &=  2 \Psi_{\textrm{B}} \left (  \frac{H}{H_{0}} \right )^{\alpha -2}           \notag \\
                                      &=  3 \Psi_{\alpha} \left (  \frac{H}{H_{0}} \right )^{\alpha -2}         ,
\label{eq:Gamma-H_power}
\end{align}
or equivalently
\begin{align}
       \Gamma      &=  3 \Psi_{\alpha} H_{0} \left (  \frac{H}{H_{0}} \right )^{\alpha -1}         ,
\label{eq:Gamma_power}
\end{align}
where  $\Psi_{\alpha} =  \frac{2}{3} \Psi_{\textrm{B}}$ given by Eq.\ (\ref{eq:Psi_a-B}) and $f_{\Lambda}(t) =0$ have also been used.
An equivalent power-law term for $\Gamma$ has been examined in CCDM models \cite{Freaza2002,Ramos_2014,Cardenas2020}.
Substituting Eq.\ (\ref{eq:Sol_HH0_power}) into Eq.\ (\ref{eq:Gamma-H_power}) yields 
\begin{align}
     \frac{  \Gamma }{ H }  &=  3 \Psi_{\alpha} \left (  \frac{H}{H_{0}} \right )^{\alpha -2}       
                                        =  \frac{ 3 \Psi_{\alpha}  } { (1- \Psi_{\alpha}) \tilde{a}^{ - \frac{3(2-\alpha)}{2}  }  + \Psi_{\alpha}  } .
\label{eq:Gamma-H_power_a}
\end{align}
When $\alpha = 2$, this equation reduces to a constant value:
\begin{align}
     \frac{  \Gamma }{ H }  &=  3 \Psi_{\alpha}  .
\label{eq:Gamma-H_power_a_alpha2}
\end{align}

Using the above equation, Eq.\ (\ref{eq:dots_s_s-a_2}) can be calculated
with $\int_{1}^{{\tilde{a}}}   \frac{ \Gamma (\tilde{a}^{\prime}) }{ H } \frac{ d\tilde{a}^{\prime} }{\tilde{a}^{\prime} }$.
Substituting Eq.\ (\ref{eq:Gamma-H_power_a}) into the integral yields
\begin{align}
\int_{1}^{{\tilde{a}}}   \frac{ \Gamma (\tilde{a}^{\prime}) }{ H } \frac{ d\tilde{a}^{\prime} }{\tilde{a}^{\prime} } 
                      &= \int_{1}^{{\tilde{a}}}   \frac{ 3 \Psi_{\alpha}   d\tilde{a}^{\prime} }{\tilde{a}^{\prime} \left [    (1- \Psi_{\alpha}) \tilde{a}^{\prime   - \frac{3(2-\alpha)}{2}  }  + \Psi_{\alpha}   \right ]  }   \notag \\
                      &= 3 \Psi_{\alpha}  \left [  \frac{  \ln  \left [   \Psi_{\alpha}   \left ( \tilde{a}^{\prime   \frac{3(2-\alpha)}{2}  }    -1 \right )     + 1      \right ]  }{  \frac{3(2-\alpha)}{2} \Psi_{\alpha}   } \right ]_{1}^{{\tilde{a}}}    \notag \\
                      &=    \ln  \left [   \Psi_{\alpha}  \tilde{a}^{\frac{3(2-\alpha)}{2}  }  +   (1 - \Psi_{\alpha} )        \right ]^{ \frac{2}{2-\alpha} }     .
\label{eq:int01}
\end{align}
Substituting Eq.\ (\ref{eq:int01}) into Eq.\ (\ref{eq:dots_s_s-a_2}) and calculating the resultant equation, we have the normalized entropy density:
\begin{align}
    \frac{s}{s_{0}}    &= \exp  \left [  \ln  \frac{ \left [   \Psi_{\alpha}  \tilde{a}^{\frac{3(2-\alpha)}{2}  }  +   (1 - \Psi_{\alpha} )        \right ]^{ \frac{2}{2-\alpha} } }{ \tilde{a}^{3} }  \right ]  \notag \\
                            &=    \tilde{a}^{-3}  \left [   \Psi_{\alpha}  \tilde{a}^{\frac{3(2-\alpha)}{2}  }  +   (1 - \Psi_{\alpha} )        \right ]^{ \frac{2}{2-\alpha} }                                  .
\label{eq:dots_s_s-a_power_0}
\end{align}
Reformulating this equation yields
\begin{align}
    \frac{s}{s_{0}}   
                            &=   \tilde{a}^{-3} \left [   \tilde{a}^{\frac{3(2-\alpha)}{2}  }  \left ( \Psi_{\alpha} +   (1 - \Psi_{\alpha} )\tilde{a}^{\frac{-3(2-\alpha)}{2}  }  \right )        \right ]^{ \frac{2}{2-\alpha} }   \notag \\
                            &=    \left [  (1 - \Psi_{\alpha} )\tilde{a}^{\frac{-3(2-\alpha)}{2}  }   +  \Psi_{\alpha}     \right ]^{ \frac{2}{2-\alpha} }   .
\label{eq:dots_s_s-a_power}
\end{align}
Equation\ (\ref{eq:dots_s_s-a_power}) can be summarized using Eq.\ (\ref{eq:Sol_HH0_power}).
The normalized entropy density relation is written as 
\begin{equation}
    \frac{ s }{ s_{0} }  = \left ( \frac{H}{H_{0}}  \right)^{2}  .
\label{eq:ss0_HH0}
\end{equation}
From this entropy density relation, the entropy in the Hubble volume is calculated in the next subsection.

\subsection{Entropy $S_{m}$ in the Hubble volume} 
\label{Entropy in the Hubble volume}

We examine the entropy $S_{m}$ in the Hubble volume and compare it with the Bekenstein--Hawking entropy $S_{\rm{BH}}$ on the Hubble horizon.
The Hubble horizon is equivalent to the apparent horizon in the flat FRW universe considered here.

The entropy $S_{m}$ in the Hubble volume is derived from the entropy density relation given by Eq.\ (\ref{eq:ss0_HH0}).
The entropy $S_{m}$ is proportional to $r_{H}^{3}$,  i.e.,  $S_{m} \propto s r_{H}^3$, where the Hubble horizon (radius) $r_{H}$ is given by 
\begin{equation}
     r_{H} = \frac{c}{H}   .
\label{eq:rH}
\end{equation}
Thus, the normalized entropy $S_{m}/S_{m,0}$ is written as
\begin{equation}
        \frac{S_{m}}{S_{m,0}} = \frac{ s r_{H}^3 }{ s_{0} r_{H0}^3}  = \left ( \frac{ s }{ s_{0} }  \right ) \left (\frac{  r_{H}^3 }{ r_{H0}^3}  \right)  ,
\label{eq:SmSm0}
\end{equation}
where $S_{m,0}$ is $S_{m}$ at the present time.
Substituting Eqs.\ (\ref{eq:ss0_HH0}) and (\ref{eq:rH}) into Eq.\ (\ref{eq:SmSm0}) yields
\begin{equation}
        \frac{S_{m}}{S_{m,0}} = \left ( \frac{H}{H_{0}}  \right)^{2}  \left (\frac{  {(c/H)}^3 }{ {(c/H_{0})}^3}  \right) =\left ( \frac{H}{H_{0}}  \right)^{-1} .
\label{eq:SmSm0-HV}
\end{equation}
The irreversible entropy $S_{m}$ in the Hubble volume is proportional to $H^{-1}$ in the present dissipative model.
This equation can be reformulated using Eq.\ (\ref{eq:Sol_HH0_power}).
Substituting Eq.\ (\ref{eq:Sol_HH0_power}) into Eq.\ (\ref{eq:SmSm0-HV}) yields
\begin{equation}  
   \frac{S_{m}}{S_{m,0}} =   \left [  (1- \Psi_{\alpha})    \tilde{a}^{ - \frac{3(2-\alpha)}{2}  }  + \Psi_{\alpha}   \right ]^{\frac{1}{\alpha-2}}  .
\label{eq:SmSm0_a-HV}
\end{equation}
The evolution of $S_{m}$ is examined using Eqs.\ (\ref{eq:SmSm0-HV}) and (\ref{eq:SmSm0_a-HV}).
We discuss this in the next section.
Note that Eq.\ (\ref{eq:SmSm0_a-HV}) reduces to $\frac{S_{m}}{S_{m,0}} =   \tilde{a}^{  \frac{3 (1- \Psi_{\alpha}) }{2}  }  $ when $\alpha = 2$.

In this section, the entropy density relation and the entropy relation, i.e., $s \propto H^{2}$ and $S_{m} \propto H^{-1}$, are derived from the present dissipative model.
The equivalent relations are obtained assuming a matter-dominated era and $s \propto a^{-3}$.
We discuss this in Appendix \ref{MDE}.

\section{Entropy evolution for the present dissipative model} 
\label{Entropy evolution}

In this section, we study the evolution of the irreversible entropy $S_{m}$ due to adiabatic particle creation in the present dissipative model.
In Sec.\ \ref{Evolution of $S_{m}$}, the second law of thermodynamics ($\dot{S}_{m} \ge 0$) and the maximization of entropy ($\ddot{S}_{m} < 0$) are discussed.
In Sec.\ \ref{Evolution of SBH}, the Bekenstein--Hawking entropy $S_{\rm{BH}}$ on the Hubble horizon is reviewed, according to a previous work \cite{Koma14}.
In Sec.\ \ref{Evolutions of Sm and SBH}, the evolutions of $S_{m}$ and $S_{\rm{BH}}$ are examined.
In Sec.\ \ref{Transition from deceleration to acceleration}, constraints on a transition from deceleration to acceleration are discussed and compared with the thermodynamic constraints.
Note that we use a normalized formulation in order to examine $S_{m}$ and $S_{\rm{BH}}$ separately.
The generalized second law and the maximization of total entropy are briefly discussed later.

\subsection{$S_{m}$, $\dot{S}_{m}$, and $\ddot{S}_{m}$ in the Hubble volume} 
\label{Evolution of $S_{m}$} 

As above, we write the irreversible entropy $S_{m}$ in the Hubble volume for the present dissipative model as
\begin{equation}  
   \frac{S_{m}}{S_{m,0}} =         \left [  (1- \Psi_{\alpha})    \tilde{a}^{ - \frac{3(2-\alpha)}{2}  }  + \Psi_{\alpha}   \right ]^{\frac{1}{\alpha -2}}     , 
\label{eq:SmSm0_a-HV_all}
\end{equation}
or equivalently, 
\begin{equation}
        \frac{S_{m}}{S_{m,0}}  =\left ( \frac{H}{H_{0}}  \right)^{-1} ,
\label{eq:SmSm0-HV2}
\end{equation}
where $H/H_{0}$ is obtained from Eq.\ (\ref{eq:Sol_HH0_power}).

\subsubsection{Second law of thermodynamics  $(\dot{S}_{m} \ge 0)$} 
\label{SL Sm} 

To discuss the second law of thermodynamics, we calculate the first derivative of $S_{m}$ for the present model.
Differentiating Eq.\ (\ref{eq:SmSm0-HV2}) with respect to $t$ and reformulating the resultant equation gives 
\begin{equation}
        \frac{\dot{S}_{m}}{ S_{m,0} H_{0}}  =   \frac{-\dot{H}}{H^{2}}  .
\label{eq:dSmSm0}
\end{equation}
Substituting Eq.\ (\ref{eq:Back_power_hB}) into Eq.\ (\ref{eq:dSmSm0})  yields
\begin{align}
        \frac{\dot{S}_{m}}{ S_{m,0} H_{0}}  &=   \frac{3}{2}  \left (  1 -   \Psi_{\alpha} \left (  \frac{H}{H_{0}} \right )^{\alpha -2} \right )    .
\label{eq:dSmSm0_all}
\end{align}
Substituting Eq.\ (\ref{eq:Sol_HH0_power}) into Eq.\ (\ref{eq:dSmSm0_all}) yields
\begin{align}
        \frac{\dot{S}_{m}}{ S_{m,0} H_{0}}  &=   \frac{3}{2}  \left (  1 -   \frac{ \Psi_{\alpha} }{  (1- \Psi_{\alpha})    \tilde{a}^{ - \frac{3(2-\alpha)}{2}  }  + \Psi_{\alpha} } \right )   \notag \\      
                                                        &=   \frac{3}{2}     \frac{ (1- \Psi_{\alpha})    \tilde{a}^{ - \frac{3(2-\alpha)}{2}  }  }{  (1- \Psi_{\alpha})    \tilde{a}^{ - \frac{3(2-\alpha)}{2}  }  + \Psi_{\alpha} }    .
\label{eq:dSmSm0_alphaNE2}
\end{align}
We can confirm that the second law of thermodynamics is satisfied, i.e., $\dot{S}_{m} \ge 0 $, because $0 \leq \Psi_{\alpha} \leq 1 $, as shown in Eq.\ (\ref{eq:Psi_01}).
Here, $H >0$ and $S_{m} >0$ have been assumed.
Of course, Eq.\ (\ref{eq:S-Gamma}) indicates the second law because $\Gamma \ge 0$ is considered.
Note that Eq.\ (\ref{eq:dSmSm0_alphaNE2}) reduces to a constant value of $\frac{3}{2} (  1 -   \Psi_{\alpha}  ) $ when $\alpha = 2$.

\subsubsection{Maximization of entropy $(\ddot{S}_{m} < 0)$} 
\label{Max Sm}

To discuss the maximization of entropy, we calculate the second derivative for the present model.
Differentiating Eq.\ (\ref{eq:dSmSm0_all}) with respect to $t$ yields
\begin{align}
     \frac{\ddot{S}_{m}}{ S_{m,0} H_{0}}  &=  \frac{d}{dt}   \left [  \frac{3}{2}  \left (  1 -   \Psi_{\alpha} \left (  \frac{H}{H_{0}} \right )^{\alpha -2} \right )   \right ]  \notag \\
&= \frac{- 3  \Psi_{\alpha}   (\alpha -2 )  }{2}  \left (  \frac{H}{H_{0}} \right  )^{\alpha -3}   \left ( \frac{ \dot{H} }{H_{0}} \right  )   .
\label{eq:d2Smdt2Sm0_cal1}
\end{align}
Reformulating Eq.\ (\ref{eq:d2Smdt2Sm0_cal1}) and substituting Eq.\ (\ref{eq:Back_power_hB}) into the resultant equation yields
\begin{align}
& \frac{\ddot{S}_{m}}{ S_{m,0} H_{0}^{2} }  = \frac{- 3  \Psi_{\alpha}  (\alpha -2 )   }{2}  \left (  \frac{H}{H_{0}} \right  )^{\alpha -1}   \left ( \frac{ \dot{H} }{H^{2}} \right  )   \notag \\
                                                        &= \frac{9    \Psi_{\alpha}   (\alpha -2 )  }{4}  \left (  \frac{H}{H_{0}} \right  )^{\alpha -1}    \left (  1 -   \Psi_{\alpha} \left (  \frac{H}{H_{0}} \right )^{\alpha -2} \right )     .
\label{eq:d2Smdt2Sm0_cal2}
\end{align}
Substituting Eq.\ (\ref{eq:Sol_HH0_power}) into Eq.\ (\ref{eq:d2Smdt2Sm0_cal2}) and calculating several operations, we obtain
\begin{align}
        \frac{\ddot{S}_{m}}{ S_{m,0} H_{0}^{2} }  
 &=  \frac{9}{4}  \frac{     (\alpha -2 ) \Psi_{\alpha} (1- \Psi_{\alpha})    \tilde{a}^{ - \frac{3(2-\alpha)}{2}  }                                                           }    
                               {    \left [  (1- \Psi_{\alpha})    \tilde{a}^{ - \frac{3(2-\alpha)}{2}  }  + \Psi_{\alpha}  \right ]^{\frac{ 3 - 2 \alpha }{ 2- \alpha }}  }  .
\label{eq:d2Smdt2Sm0_alphaNE2}
\end{align}
The above equation indicates that $ \ddot{S}_{m} < 0  $ is satisfied when $\alpha < 2$.
When $\alpha = 2$, this equation reduces to $0$.
Accordingly, to satisfy the maximization of entropy, we require
\begin{align}  
  \alpha < 2    ,
\label{eq:alpha_d2Smdt2}
\end{align}
where $ 0 < \Psi_{\alpha} < 1 $ is assumed.
In addition, Eq.\ (\ref{eq:d2Smdt2Sm0_alphaNE2}) implies that $\ddot{S}_{m}$ for all $\alpha$ approaches $0$ in the last stage ($1 \ll \tilde{a}$).
We discuss this in Sec.\ \ref{Evolutions of Sm and SBH}.
In the next subsection, the Bekenstein--Hawking entropy is examined in the present dissipative model.

\subsection{$S_{\rm{BH}}$, $\dot{S}_{\rm{BH}}$, and $\ddot{S}_{\rm{BH}}$ on the Hubble horizon} 
\label{Evolution of SBH} 

We assume that the horizon of the universe has an associated entropy, i.e., the Bekenstein--Hawking entropy, extending the concept of black hole thermodynamics \cite{Easson,Cai,Basilakos1,Koma4,Koma5}.
The Bekenstein--Hawking entropy $S_{\rm{BH}}$ is written as 
\begin{equation}
 S_{\rm{BH}}  = \frac{ k_{B} c^3 }{  \hbar G }  \frac{A_{H}}{4}   ,
\label{eq:SBH}
\end{equation}
where $\hbar$ is the reduced Planck constant defined as $\hbar \equiv h/(2 \pi)$, using the Planck constant $h$ \cite{Koma1012,Koma11,Koma14}. 
$A_{H}$ is the surface area of the sphere with the Hubble horizon $r_{H}$.
In a flat FRW universe, the Hubble horizon is equivalent to the apparent horizon.
Substituting $A_{H}=4 \pi r_{H}^2 $ into Eq.\ (\ref{eq:SBH}) and applying Eq.\ (\ref{eq:rH}) yields 
\begin{equation}
S_{\rm{BH}} =  \left ( \frac{ \pi k_{B} c^5 }{ \hbar G } \right )  \frac{1}{H^2}    =    \frac{K}{H^2}   ,
\label{eq:SBH2}      
\end{equation}
where $K$ is a positive constant given by \cite{Koma4,Koma5}
\begin{equation}
  K =  \frac{  \pi  k_{B}  c^5 }{ \hbar G } .
\label{eq:K-def}
\end{equation}
Using $S_{\rm{BH},0} = K/H_{0}^{2}$, the normalized Bekenstein--Hawking entropy is written as
\begin{equation}
\frac{S_{\rm{BH}}}{ S_{\rm{BH},0} } =  \left ( \frac{ H }{ H_{0} } \right )^{-2}  , 
\label{eq:SBHSBH0_0}      
\end{equation}
where $S_{\rm{BH},0}$ is the Bekenstein--Hawking entropy at the present time.
Equation\ (\ref{eq:SBHSBH0_0}) indicates that the normalized Bekenstein--Hawking entropy depends on the background evolution of the universe and is generally proportional to $(H/H_{0})^{-2}$.
Note that cosmological models have not yet been assumed in the above discussion.

\subsubsection{$S_{\rm{BH}}$, $\dot{S}_{\rm{BH}}$, and $\ddot{S}_{\rm{BH}}$ for the present dissipative model}
\label{SBH dSBH d2SBH for the present dissipative model}

We now discuss $S_{\rm{BH}}$, $\dot{S}_{\rm{BH}}$, and $\ddot{S}_{\rm{BH}}$ for the present dissipative model.
As mentioned in Sec.\ \ref{BV model with a power-law term}, the background evolution of the dissipative universe considered here is equivalent to that of a non-dissipative universe examined in a previous work \cite{Koma14}. 
Therefore, $S_{\rm{BH}}$ in Ref.\ \cite{Koma14} can be applied to the present model because $S_{\rm{BH}}$ depends on the background evolution.
The result is summarized in Appendix \ref{Entropy on the horizon}.
From Eq. (\ref{eq:SBHSBH0_power}), the normalized $S_{\rm{BH}}$ is written as  
\begin{equation}  
   \frac{ S_{\rm{BH}}  }{S_{\rm{BH},0}  } =   \left (     (1- \Psi_{\alpha})   \tilde{a}^{ - \frac{3(2-\alpha)}{2}  }  + \Psi_{\alpha}     \right )^{\frac{2}{\alpha-2}}        .
\label{eq:SBHSBH0_power_0}
\end{equation}
From Eq. (\ref{eq:dSBHSBH0_power}), the normalized $\dot{S}_{\rm{BH}}$ is written as 
\begin{equation}  
   \frac{ \dot{S}_{\rm{BH}}  }{S_{\rm{BH},0} H_{0} } =    \frac{ 3 (1- \Psi_{\alpha})   \tilde{a}^{ - \frac{3(2-\alpha)}{2}  }   }{  \left [ (1- \Psi_{\alpha})   \tilde{a}^{ - \frac{3(2-\alpha)}{2}  }  + \Psi_{\alpha}  \right ]^{ \frac{3- \alpha}{2-\alpha} }  }      .  
\label{eq:dSBHSBH0_power_0}
\end{equation}
This equation indicates that the present model always satisfies $\dot{S}_{\rm{BH}} \ge 0$, because $ 0 \leq \Psi_{\alpha} \leq 1 $ is assumed \cite{Koma14}.
In addition, from Eq. (\ref{eq:d2SBH2SBH0_power}), the normalized $\ddot{S}_{\rm{BH}}$ is 
\begin{equation}  
   \frac{ \ddot{S}_{\rm{BH}}  }{S_{\rm{BH},0} H_{0}^{2} } =    
                                                               \frac{9}{2}    \frac{  (1- \Psi_{\alpha})    \tilde{a}^{ - \beta }     \left [  (1- \Psi_{\alpha})    \tilde{a}^{ - \beta }    + (\alpha-2)\Psi_{\alpha}   \right ]             }{   \left [    (1- \Psi_{\alpha})   \tilde{a}^{ - \beta  }  + \Psi_{\alpha}         \right ]^{2}      }        ,  
\label{eq:d2SBH2SBH0_power_0}
\end{equation}
where a parameter $\beta$ is used for simplicity, given by
\begin{equation}  
                                 \beta  = \frac{3(2-\alpha)}{2} .
\label{eq:beta_0}
\end{equation}
Equation\ (\ref{eq:d2SBH2SBH0_power_0}) is slightly complicated.
In fact, this equation indicates that $ \ddot{S}_{\rm{BH}} < 0$ should be satisfied at least in the last stage, i.e., $ \tilde{a} \rightarrow \infty$, when $\alpha < 2$ \cite{Koma14}.

In this way, we can obtain the three parameters $S_{\rm{BH}}$, $\dot{S}_{\rm{BH}}$, and $\ddot{S}_{\rm{BH}}$, for the present dissipative universe.
For details, see Appendix \ref{Entropy on the horizon}.

\begin{figure*} [htb]  
 \begin{minipage}{0.495\hsize}
  \begin{center}
   \scalebox{0.3}{\includegraphics{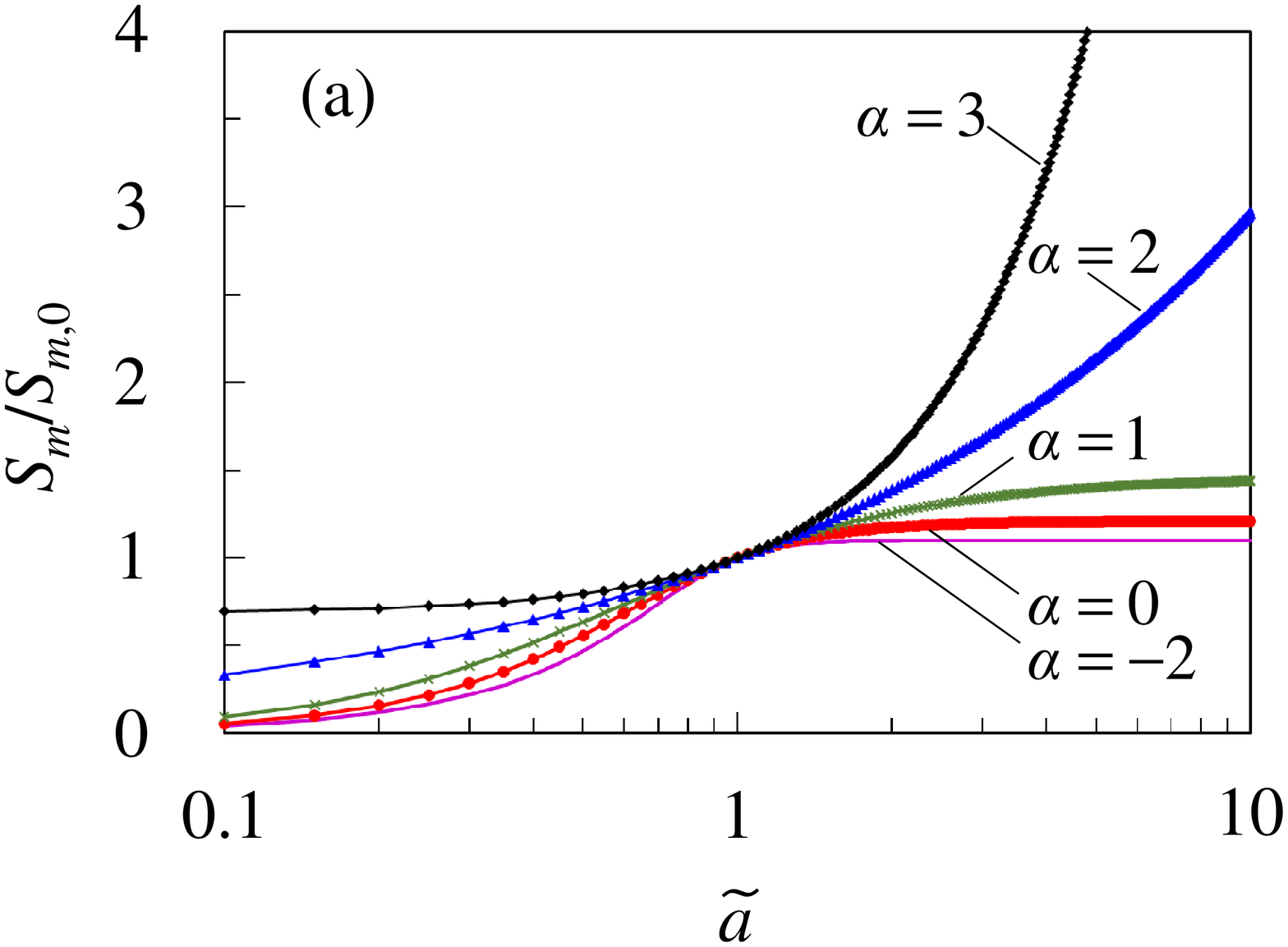}}\\    
  \end{center}
 \end{minipage}
 \begin{minipage}{0.495\hsize}
  \begin{center}
   \scalebox{0.3}{\includegraphics{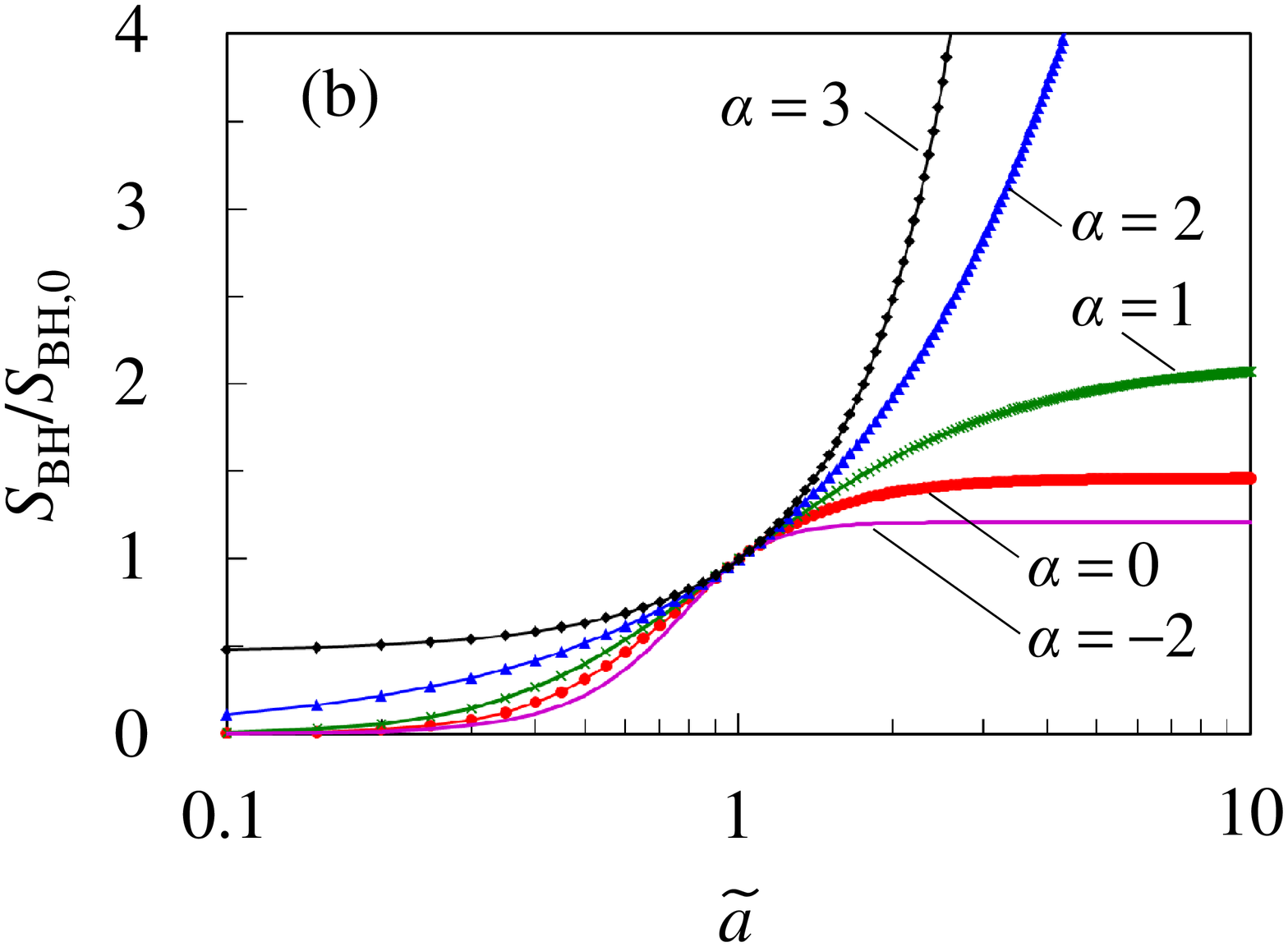}}\\
  \end{center}
 \end{minipage}
\caption{ (Color online). Evolutions of $S_{m}$ and $S_{\rm{BH}}$ for $\Psi_{\alpha} =0.685$. (a) Normalized $S_{m}$.  (b) Normalized $S_{\rm{BH}}$.  
The background evolution of the dissipative universe is set to be equivalent to that of a non-dissipative universe, examined in Ref.\ \cite{Koma14}.
Therefore, $S_{\rm{BH}}$ in (b), $\dot{S}_{\rm{BH}}$ in Fig.\ \ref{Fig-dSm_dSBH}(b), and $\ddot{S}_{\rm{BH}}$ in Fig.\ \ref{Fig-d2Sm_d2SBH}(b) are essentially equivalent to those in Ref.\ \cite{Koma14}. 
However, in (a), irreversible entropy $S_{m}$ is produced in the dissipative universe, unlike in the non-dissipative universe.}
\label{Fig-Sm_SBH}
\end{figure*}

\begin{figure*} [htb] 
 \begin{minipage}{0.495\hsize}
  \begin{center}
   \scalebox{0.3}{\includegraphics{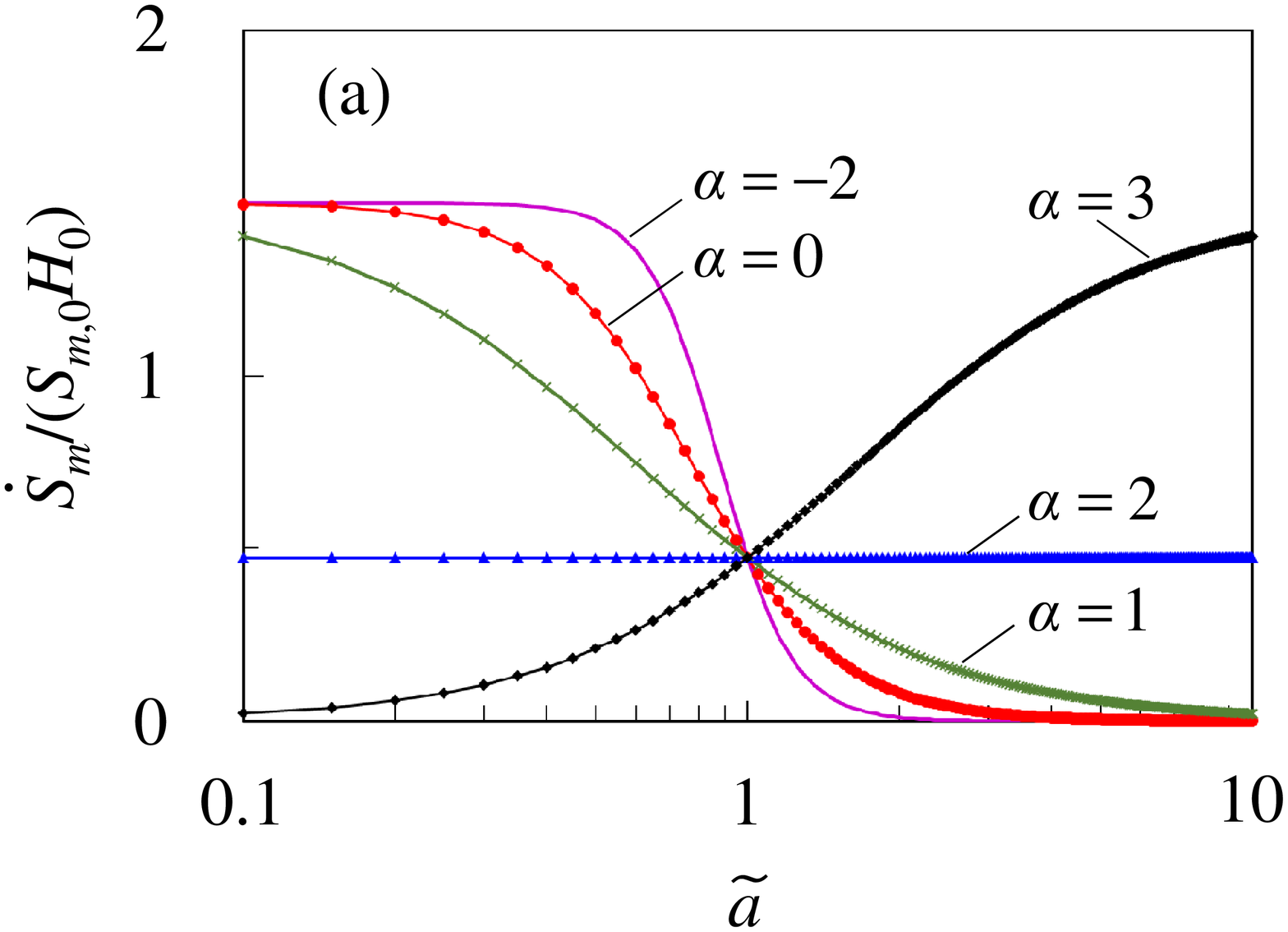}}\\  
  \end{center}
 \end{minipage}
 \begin{minipage}{0.495\hsize}
  \begin{center}
   \scalebox{0.3}{\includegraphics{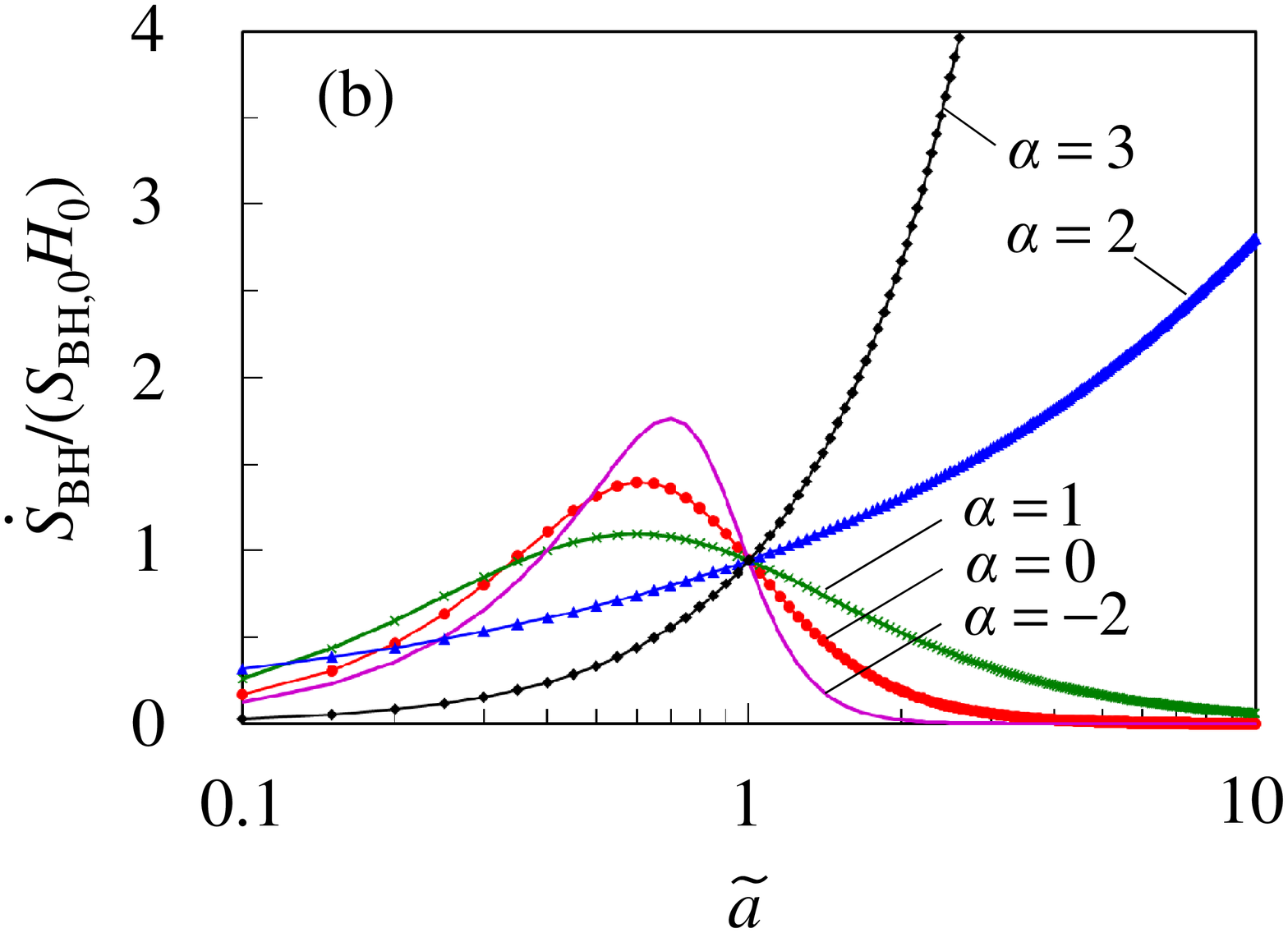}}\\
  \end{center}
 \end{minipage}
\caption{ (Color online). Evolutions of $\dot{S}_{m}$ and $\dot{S}_{\rm{BH}}$ for $\Psi_{\alpha} =0.685$.  (a) Normalized $\dot{S}_{m}$. (b) Normalized $\dot{S}_{\rm{BH}}$.
For $\dot{S}_{\rm{BH}}$, see the caption of Fig.\ \ref{Fig-Sm_SBH}. }
\label{Fig-dSm_dSBH}
\end{figure*}

\begin{figure*} [htb] 
 \begin{minipage}{0.495\hsize}
  \begin{center}
   \scalebox{0.3}{\includegraphics{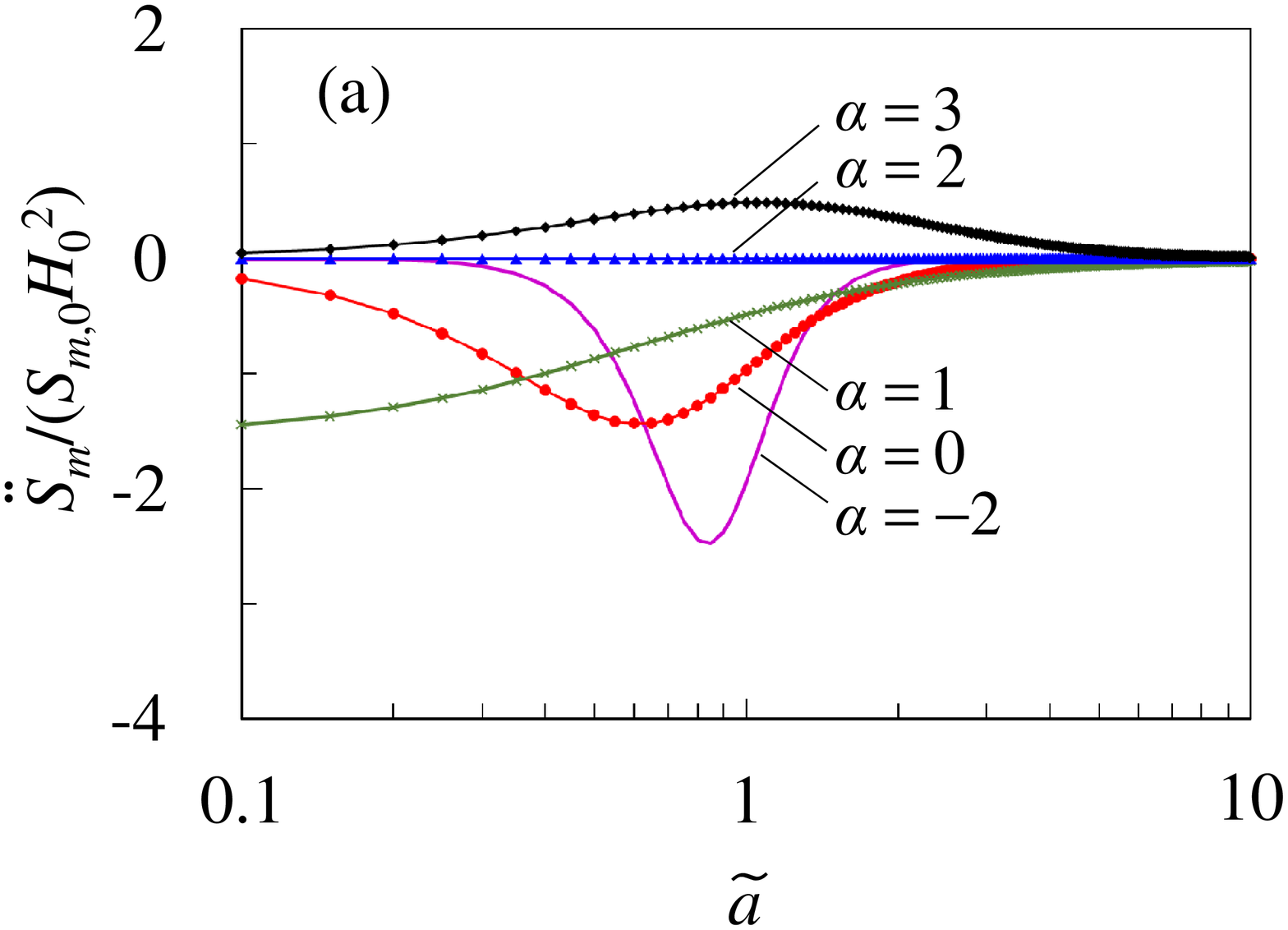}}\\
  \end{center}
 \end{minipage}
 \begin{minipage}{0.495\hsize}
  \begin{center}
   \scalebox{0.3}{\includegraphics{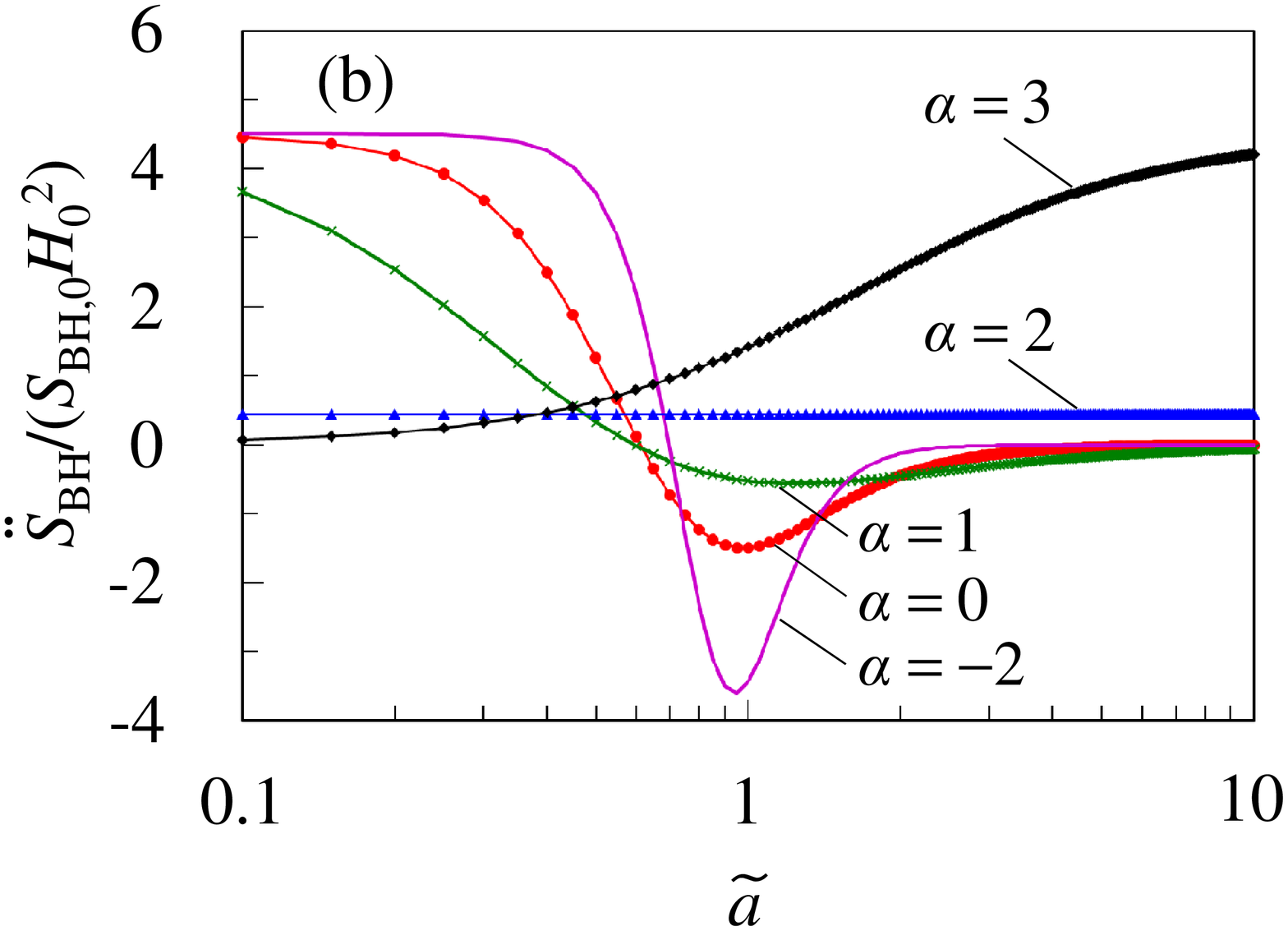}}\\
  \end{center}
 \end{minipage}
\caption{ (Color online). Evolutions of $\ddot{S}_{m}$ and $\ddot{S}_{\rm{BH}}$ for $\Psi_{\alpha} =0.685$. (a) Normalized $\ddot{S}_{m}$. (b) Normalized $\ddot{S}_{\rm{BH}}$.
The normalized $\ddot{S}_{m}$ for $\alpha =2$ is $0$ from Eq.\ (\ref{eq:d2Smdt2Sm0_alphaNE2}), whereas the normalized $\ddot{S}_{\rm{BH}}$ for $\alpha =2$ is approximately $0.447$ from Eq.\ (\ref{eq:d2SBH2SBH0_power_0}). 
For $\ddot{S}_{\rm{BH}}$, see the caption of Fig.\ \ref{Fig-Sm_SBH}. }
\label{Fig-d2Sm_d2SBH}
\end{figure*}

\begin{figure*} [htb] 
 \begin{minipage}{0.495\hsize}
  \begin{center}
   \scalebox{0.3}{\includegraphics{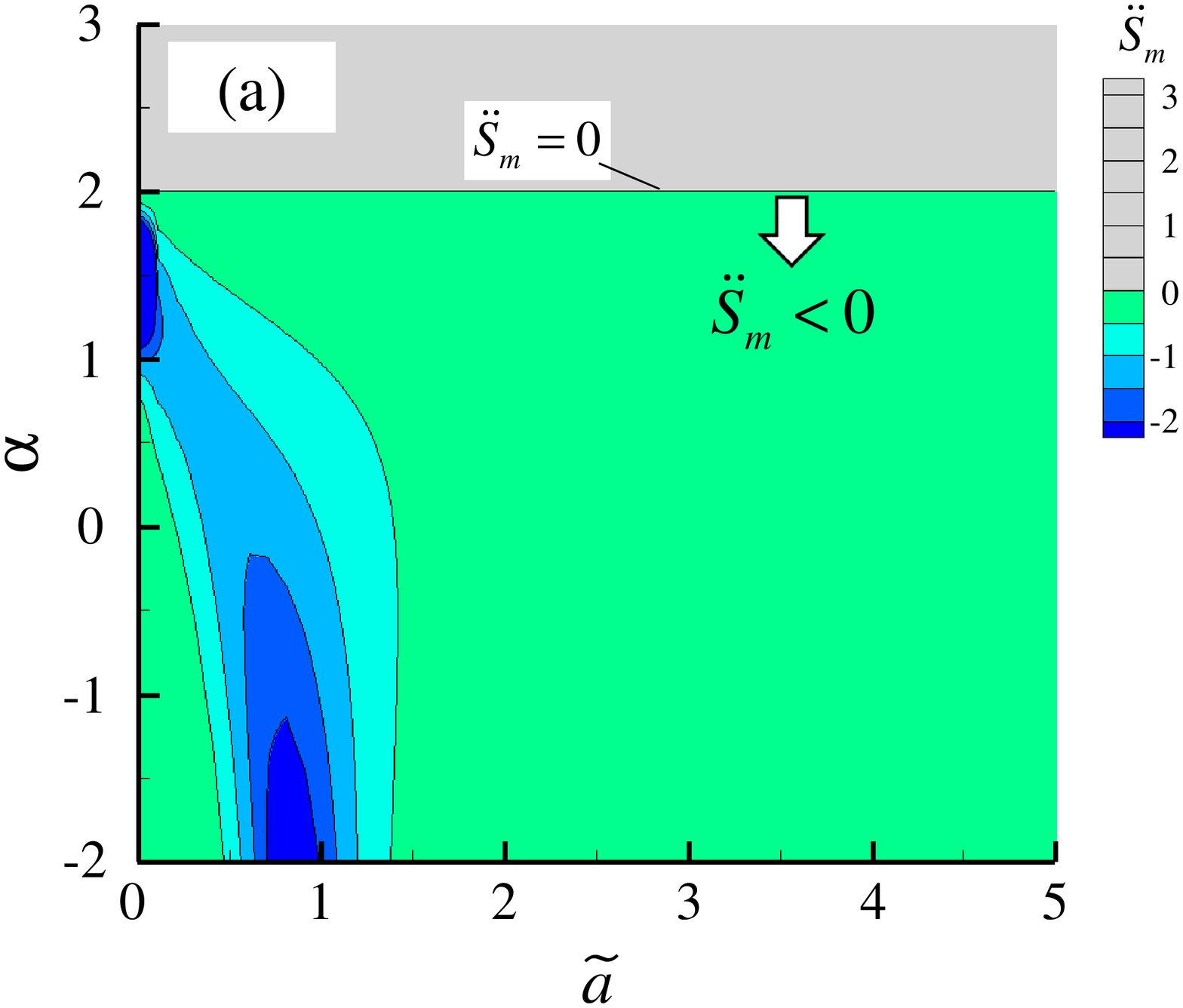}}\\  
  \end{center}
 \end{minipage}
 \begin{minipage}{0.495\hsize}
  \begin{center}
   \scalebox{0.3}{\includegraphics{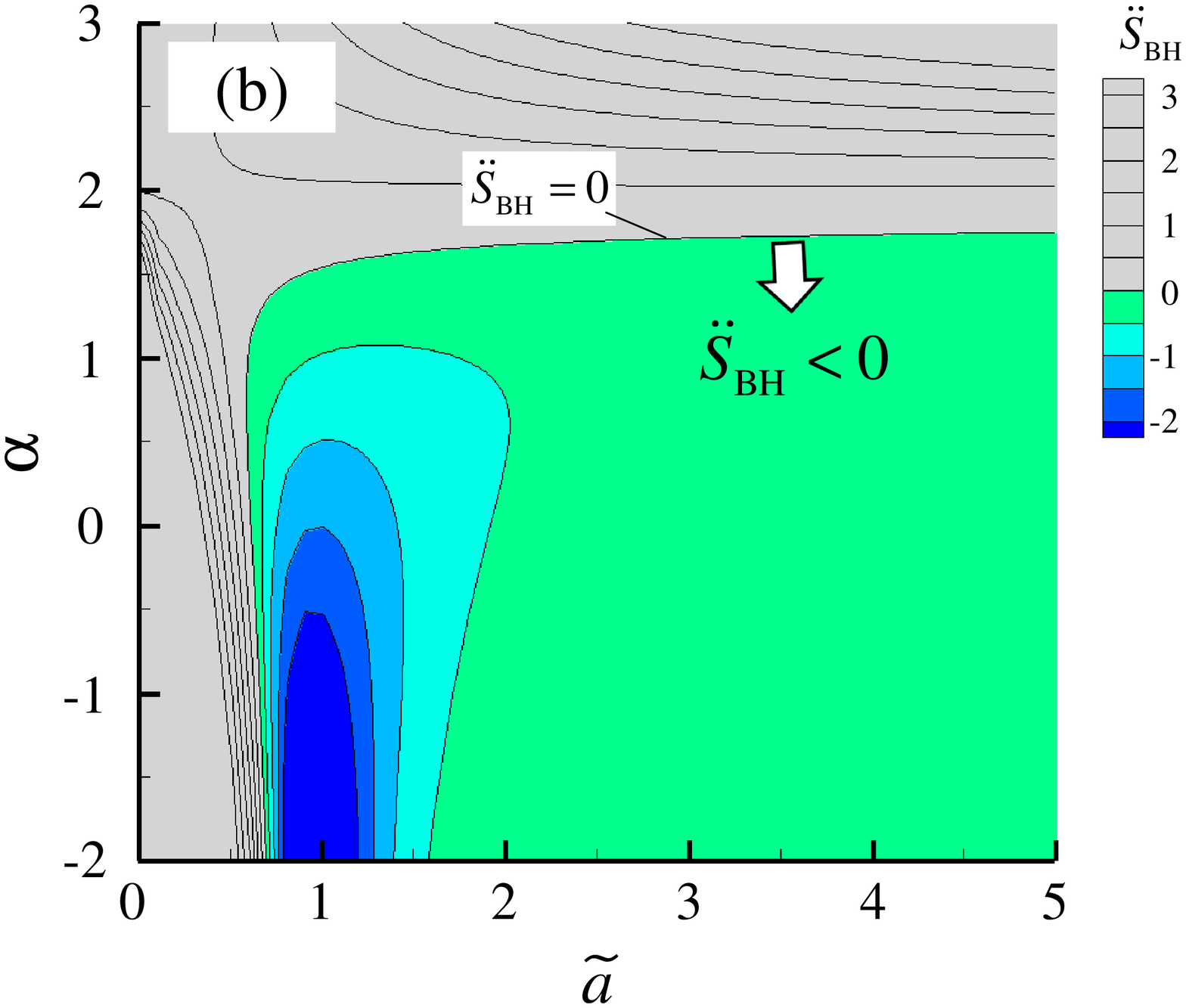}}\\
  \end{center}
 \end{minipage}
\caption{ (Color online). Contours of $\ddot{S}_{m}$ and $\ddot{S}_{\rm{BH}}$ in the $(\tilde{a}, \alpha)$ plane for $\Psi_{\alpha} =0.685$.
(a) Normalized $\ddot{S}_{m}$. (b) Normalized $\ddot{S}_{\rm{BH}}$.
The arrow indicates a region that satisfies $\ddot{S}_{m} < 0$ in (a) and $\ddot{S}_{\rm{BH}} < 0$ in (b).
Unsatisfied regions are displayed in gray, to make the boundary of $\ddot{S}_{m} = 0$ in (a) and $\ddot{S}_{\rm{BH}} = 0$ in (b) clear. 
The color scale bar is based on the normalized value.
The contour lines are plotted at increments of $0.5$. }
\label{Fig-d2Sm_d2SBH_map}
\end{figure*}

\begin{figure*} [htb] 
 \begin{minipage}{0.495\hsize}
  \begin{center}
   \scalebox{0.3}{\includegraphics{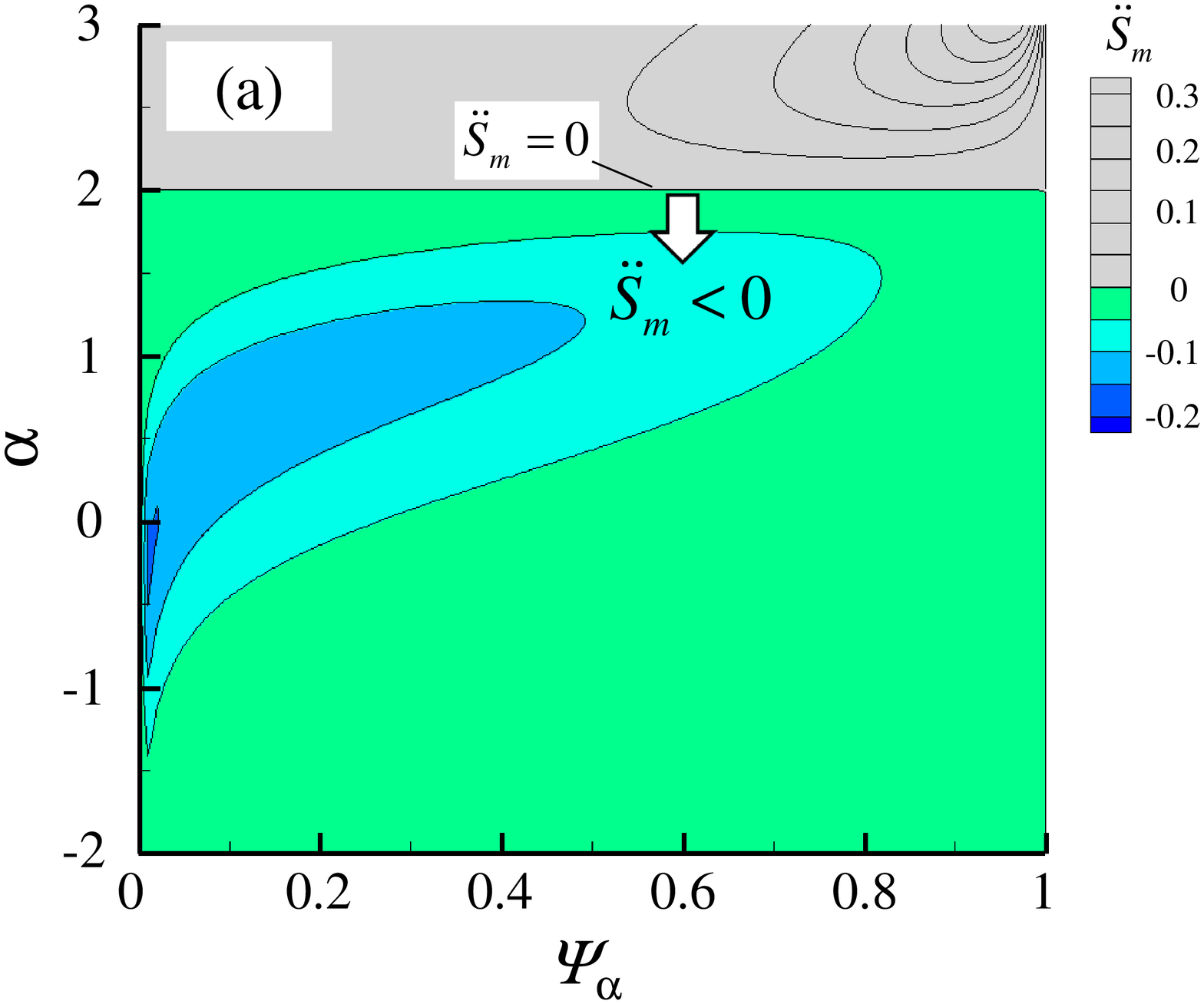}}\\  
  \end{center}
 \end{minipage}
 \begin{minipage}{0.495\hsize}
  \begin{center}
   \scalebox{0.3}{\includegraphics{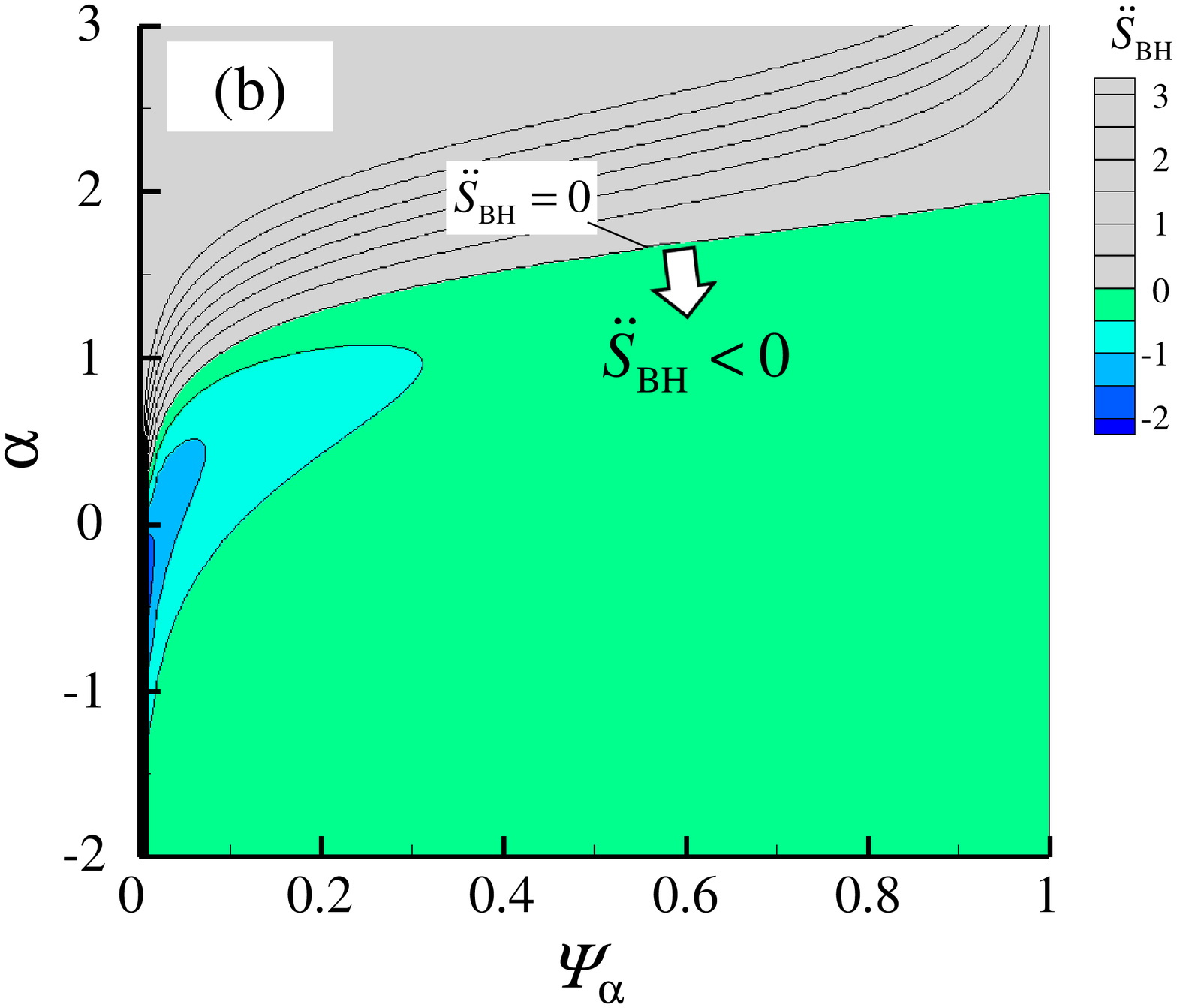}}\\
  \end{center}
 \end{minipage}
\caption{ (Color online). Contours of $\ddot{S}_{m}$ and $\ddot{S}_{\rm{BH}}$ in the $(\Psi_{\alpha}, \alpha)$ plane for $\tilde{a}=5$. 
(a) Normalized $\ddot{S}_{m}$.  (b) Normalized $\ddot{S}_{\rm{BH}}$.
The horizontal axis represents $\Psi_{\alpha}$, which is a type of density parameter for the effective dark energy.
The arrow indicates a region that satisfies $\ddot{S}_{m} < 0$ in (a) and $\ddot{S}_{\rm{BH}} < 0$ in (b).
Unsatisfied regions are displayed in gray, to make the boundary of $\ddot{S}_{m} = 0$ in (a) and $\ddot{S}_{\rm{BH}} = 0$ in (b) clear. 
The color scale bar is based on the normalized value.
In (a), the contour lines are plotted at increments of $0.05$, and in (b) at increments of $0.5$. }
\label{Fig-d2Sm_d2SBH_map_psi-alpha}
\end{figure*}

\subsection{Evolutions of $S_{m}$ and $S_{\rm{BH}}$} 
\label{Evolutions of Sm and SBH} 

In this subsection, we examine the evolution of the irreversible entropy $S_{m}$ and the Bekenstein--Hawking entropy $S_{\rm{BH}}$ for the present dissipative model.

Figures\ \ref{Fig-Sm_SBH}, \ref{Fig-dSm_dSBH}, and \ref{Fig-d2Sm_d2SBH} show the evolutions of the entropies ($S_{m}$ and $S_{\rm{BH}}$), the first derivatives ($\dot{S}_{m}$ and $\dot{S}_{\rm{BH}}$), and the second derivatives ($\ddot{S}_{m}$ and $\ddot{S}_{\rm{BH}}$), respectively.
The horizontal axis represents the normalized scale factor, $ \tilde{a} =a/a_{0}$, where $\tilde{a}$ increases with time because an expanding universe is assumed.
In these figures, $\alpha$ is set to $-2$, $0$, $1$, $2$, and $3$, to show typical results.
Also, $\Psi_{\alpha}$ is set to $0.685$, as examined in Fig.\ \ref{Fig-H-a}.
The background evolution of the dissipative universe is equivalent to that of the non-dissipative universe examined in a previous work \cite{Koma14}
and the evolutions of $S_{\rm{BH}}$, $\dot{S}_{\rm{BH}}$, and $\ddot{S}_{\rm{BH}}$ are essentially equivalent to those in Ref.\ \cite{Koma14}.
However, an irreversible entropy $S_{m}$ is produced in the dissipative universe, unlike in the non-dissipative universe [Figs.\ \ref{Fig-Sm_SBH}(a), \ref{Fig-dSm_dSBH}(a), and \ref{Fig-d2Sm_d2SBH}(a)].

As shown in Fig.\ \ref{Fig-Sm_SBH}(a), for all $\alpha$, the normalized $S_{m}$ increases with $\tilde{a}$.
Similarly, the normalized $S_{\rm{BH}}$ increases with $\tilde{a}$ [Fig.\ \ref{Fig-Sm_SBH}(b)].
Therefore, both the normalized $\dot{S}_{m}$ and the normalized $\dot{S}_{\rm{BH}}$ are non-negative [Fig.\ \ref{Fig-dSm_dSBH}].
That is, the second law of thermodynamics is satisfied for both $S_{m}$ and $S_{\rm{BH}}$.
Accordingly, the generalized second law, i.e., $\dot{S}_{m} + \dot{S}_{\rm{BH}} \geq 0$, is also satisfied.
However, the evolution of $\dot{S}_{m}$ is different from that of $\dot{S}_{\rm{BH}}$.
For example, the normalized $\dot{S}_{m}$ for $\alpha <2$ decreases with $\tilde{a}$ [Fig.\ \ref{Fig-dSm_dSBH}(a)],
while the normalized $\dot{S}_{\rm{BH}}$ for $\alpha <2$ increases with $\tilde{a}$ in the early stage and thereafter gradually decreases with $\tilde{a}$ [Fig.\ \ref{Fig-dSm_dSBH}(b)].
In addition, when $\alpha =3$, the normalized $\dot{S}_{m}$ increases slowly in the last stage, whereas $\dot{S}_{\rm{BH}}$ increases rapidly.

Consequently, the normalized $\ddot{S}_{m}$ for $\alpha < 2$ is always negative (and $\ddot{S}_{m}$ for $\alpha = 2$ is zero), as shown in Fig.\ \ref{Fig-d2Sm_d2SBH}(a).
Therefore, maximization of entropy for $S_{m}$,  $ \ddot{S}_{m} < 0  $, is always satisfied when $\alpha < 2$.
In addition, the normalized $\ddot{S}_{m}$ for $\alpha = 3$ gradually approaches zero although it is positive in the early stage.
In fact, the normalized $\ddot{S}_{m}$ for all $\alpha$ finally approaches $0$ in the last stage.

In contrast, the normalized $\ddot{S}_{\rm{BH}}$ for $\alpha < 2$ is positive in the early stage and negative in the last stage [Fig.\ \ref{Fig-d2Sm_d2SBH}(b)].
When $\alpha = 3$, the normalized $\ddot{S}_{\rm{BH}}$ is positive and increases with $\tilde{a}$.
Accordingly, maximization of entropy for $S_{\rm{BH}}$, i.e., $\ddot{S}_{\rm{BH}} <0$, is not satisfied when $\alpha \ge 2$, but should be satisfied at least in the last stage when $\alpha <2$.
The result for $\ddot{S}_{\rm{BH}}$ is consistent with that for a non-dissipative universe \cite{Koma14}.

As observed above, when $\alpha < 2$, $ \ddot{S}_{m} < 0$ is always satisfied, whereas $\ddot{S}_{\rm{BH}} <0$ should not be satisfied in the early stage.
Accordingly, we systematically examine the evolution of an $\alpha$-region that satisfies the maximization of entropy. 
To this end, we plot contours of $\ddot{S}_{m}$ and $\ddot{S}_{\rm{BH}}$ in the $(\tilde{a}, \alpha)$ plane.
As shown in Fig.\ \ref{Fig-d2Sm_d2SBH_map}, the horizontal axis represents the normalized scale factor $\tilde{a}$, which increases with time.
The vertical axis represents a parameter $\alpha$, which is used as a power-law term proportional to $H^{\alpha}$.
The arrow indicates a region that satisfies the maximization of entropy, $\ddot{S}_{m} < 0$ in Fig.\ \ref{Fig-d2Sm_d2SBH_map}(a) and $\ddot{S}_{\rm{BH}} < 0$ in Fig.\ \ref{Fig-d2Sm_d2SBH_map}(b).
Figure\ \ref{Fig-d2Sm_d2SBH_map} includes plots shown in Fig.\ \ref{Fig-d2Sm_d2SBH}.

As shown in Fig.\ \ref{Fig-d2Sm_d2SBH_map}(a), the normalized $\ddot{S}_{m}$ is always negative (in the early and last stages) when $\alpha < 2$.
In contrast, even when $\alpha < 2$, the normalized $\ddot{S}_{\rm{BH}}$ is positive in the early stage ($\tilde{a} \ll 1$) and should be negative in the last stage [Fig.\ \ref{Fig-d2Sm_d2SBH_map}(b)].
The two results indicate that constraints on $\ddot{S}_{\rm{BH}}<0$ are slightly tighter than those on $\ddot{S}_{m}<0$.

In the above discussion, we have set $\Psi_{\alpha} = 0.685$.
In the present model, $\Psi_{\alpha}$ is a type of density parameter for effective dark energy.
To examine the effect of $\Psi_{\alpha}$, we plot contours of $\ddot{S}_{m}$ and $\ddot{S}_{\rm{BH}}$ in the $(\Psi_{\alpha}, \alpha)$ plane.
In Fig.\ \ref{Fig-d2Sm_d2SBH_map_psi-alpha}, $\tilde{a}$ is set to $5$, corresponding to the last stage shown in Fig.\ \ref{Fig-d2Sm_d2SBH_map}.

As shown in Fig.\ \ref{Fig-d2Sm_d2SBH_map_psi-alpha}(a), the normalized $\ddot{S}_{m}$ is negative when $\alpha < 2$.
In contrast, when $\alpha < 2$, $\ddot{S}_{\rm{BH}} < 0$ is almost satisfied, except for a small-$\Psi_{\alpha}$ and large-$\alpha$ region [Fig.\ \ref{Fig-d2Sm_d2SBH_map_psi-alpha}(b)].
Therefore, the maximization of entropy for $S_{\rm{BH}}$, $\ddot{S}_{\rm{BH}} < 0$, has not yet been satisfied in the small-$\Psi_{\alpha}$ and large-$\alpha$ region.
It should take a long time to satisfy $\ddot{S}_{\rm{BH}} < 0$ in this region, even when $\alpha <2$, as discussed in Ref.\ \cite{Koma14}.

In this way, the conditions for satisfying $\ddot{S}_{\rm{BH}} <0$ are tighter than those for $\ddot{S}_{m} <0$ in the dissipative universe.
So far, we have discussed $\ddot{S}_{m}$ and $\ddot{S}_{\rm{BH}}$ separately.
Finally, we consider conditions to satisfy the maximization of total entropy, i.e., $\ddot{S}_{m} + \ddot{S}_{\rm{BH}} <0$.
For this, the second derivative itself should be discussed.
As shown in Figs.\ \ref{Fig-d2Sm_d2SBH_map} and \ref{Fig-d2Sm_d2SBH_map_psi-alpha}, the order of the normalized $|\ddot{S}_{\rm{BH}}|$ is approximately the same as that of the normalized $|\ddot{S}_{m}|$.
Of course, it is well-known that the horizon entropy is extremely large compared to the other entropies \cite{Egan1}.
Accordingly, $|\ddot{S}_{\rm{BH}}|$ is larger than $|\ddot{S}_{m}|$.
In addition, as noted above the conditions for satisfying $\ddot{S}_{\rm{BH}} <0$ are tighter than those for $\ddot{S}_{m} <0$.
From these two results, we can expect that the conditions for satisfying $\ddot{S}_{m} + \ddot{S}_{\rm{BH}} <0$ depend almost entirely on the conditions for satisfying $\ddot{S}_{\rm{BH}} <0$.
Consequently, the maximization of total entropy should be satisfied at least in the last stage when $\alpha < 2$.

More detailed calculations are required if the conditions for satisfying $\ddot{S}_{m} <0$ are tighter than those for $\ddot{S}_{\rm{BH}} <0$, unlike for the present dissipative universe.
However, the latter is tighter than the former in the present model and
therefore, as discussed in the above paragraph, we can reach the approximate conclusion without using a detailed calculation.
For detailed calculations, see, e.g., the work of Sol\`{a} and Yu \cite{Sola2020}.

\subsection{Transition from deceleration to acceleration}
\label{Transition from deceleration to acceleration} 

As examined in Sec.\ \ref{Evolutions of Sm and SBH}, the thermodynamic constraints on the present dissipative model should be $\alpha < 2$, at least in the last stage.
Similarly, observations are expected to constrain the present model. 
For example, observations imply an initially decelerating and then accelerating universe \cite{PERL1998_Riess1998,Planck2018}.
Accordingly, in this subsection, we study constraints on the transition and compare them with the thermodynamic constraints.

\begin{figure} [t] 
\begin{minipage}{0.495\textwidth}
\begin{center}
\scalebox{0.3}{\includegraphics{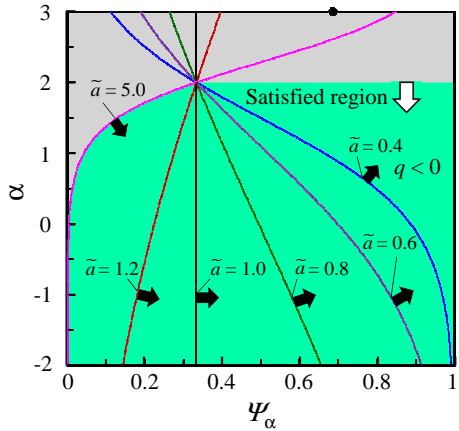}}
\end{center}
\end{minipage}
\caption{ (Color online).  Boundary of $q= 0$ in the $(\Psi_{\alpha}, \alpha)$ plane for various values of $\tilde{a}$ and a satisfied region for $\tilde{a}=5$.
The white arrow indicates a region that satisfies a transition from deceleration to acceleration when $\tilde{a}=5$.
The satisfied region for $\tilde{a}=5$ is displayed in green and the unsatisfied region is displayed in gray.
The boundaries for $\tilde{a}=0.4$, $0.6$, $0.8$, $1.0$, $1.2$, and $5.0$ are also shown.
The black arrow on each boundary indicates an accelerating-universe region that satisfies $q< 0$.
The closed circle represents $(\Psi_{\alpha},\alpha) = (0.685, 3)$, corresponding to the plot for $\alpha =3$ shown in Fig.\ \ref{Fig-H-a}(b).
The intersection point is $(\Psi_{\alpha}, \alpha) = (\frac{1}{3}, 2)$.
A similar boundary of $q= 0$ has been examined for a non-dissipative universe \cite{Koma14}. }
\label{Fig-q_plane_power}
\end{figure}

To examine a transition from deceleration to acceleration, we use the boundary required for $q = 0$ given by Eq.\ (\ref{eq:q=0}).
Figure\ \ref{Fig-q_plane_power} shows the boundary for $q= 0$ in the $(\Psi_{\alpha}, \alpha)$ plane for various values of $\tilde{a}$ and a `satisfied region' for $\tilde{a}=5$.
The white arrow indicates the region that satisfies the transition from deceleration to acceleration when $\tilde{a}=5$.
In this figure, $\tilde{a}$ is set to $0.4$, $0.6$, $0.8$, $1.0$, $1.2$, and $5.0$, to examine typical boundaries in the past and future.
The black arrow on each boundary indicates an accelerating-universe region that satisfies $q< 0$.
A similar boundary has been examined in a non-dissipative universe \cite{Koma14}.
Note that the accelerating-universe region for $\tilde{a}=5$ is different from the satisfied region for $\tilde{a}=5$.

As shown in Fig.\ \ref{Fig-q_plane_power}, the accelerating-universe region varies with $\tilde{a}$, although the intersection point $(\Psi_{\alpha}, \alpha) = (\frac{1}{3}, 2)$ is fixed \cite{Koma14}.
For example, the boundary for $\tilde{a}=0.4$ indicates that large values of $\Psi_{\alpha}$ and $\alpha$ tend to yield an accelerating universe.
When $\alpha <2$, the accelerating-universe region gradually extends with increasing $\tilde{a}$.
Accordingly, $\alpha <2$ should correspond to a decelerating and accelerating universe, at least in the last stage.

In contrast, when $\alpha >2$, the decelerating-universe region gradually extends with increasing $\tilde{a}$. 
Therefore, an accelerating and decelerating universe is expected \cite{Koma14}, as shown in Fig.\ \ref{Fig-H-a}(b).
To examine this, we focus on the point $(\Psi_{\alpha}, \alpha) = (0.685, 3)$, corresponding to the plot for $\alpha =3$ shown in Fig.\ \ref{Fig-H-a}(b).
The boundaries for $\tilde{a} =0.4$--$1.2$ indicate that the point $(0.685, 3)$ is inside the accelerating-universe region [Fig.\ \ref{Fig-q_plane_power}].
However, the boundary for $\tilde{a} =5.0$ indicates that the point is outside the region, i.e., it is inside the decelerating-universe region.
Therefore, $\alpha >2$ corresponds to an accelerating and decelerating universe \cite{Koma14}.
Consequently, $\alpha >2$ is not included in the satisfied region for $\tilde{a}=5$.

These results indicate that a transition from deceleration to acceleration is not satisfied when $\alpha >2$.
In contrast, when $\alpha <2$, the transition should be satisfied at least in the last stage.
This constraint, i.e., $\alpha <2$, is likely consistent with the thermodynamic constraint examined in Sec.\ \ref{Evolutions of Sm and SBH}.
Of course, the constraint on the transition becomes tighter when the transition point is set to be, e.g., $\tilde{a} =0.6$.
In this sense, the constraint on the transition is tighter than for $\alpha <2$, as for constraints on $\ddot{S}_{\rm{BH}} <0$.
Detailed studies are left for future research.

It should be noted that constraints on a transition from deceleration to acceleration can be considered by focusing on $\tilde{a} \rightarrow 0$ and $\tilde{a} \rightarrow \infty$.
For simplicity, $0 < \Psi_{\alpha} < 1$ is assumed here.
When $\tilde{a} \rightarrow 0$, Eq.\ (\ref{eq:q_power}) reduces to $q \approx 1/2$ for $\alpha < 2$ and $q \approx -1$ for $\alpha > 2$.
Therefore, in the early stage, $\alpha < 2$ corresponds to deceleration, and $\alpha > 2$ corresponds to acceleration.
In contrast, when $\tilde{a} \rightarrow \infty$, Eq.\ (\ref{eq:q_power}) reduces to $q \approx -1$ for $\alpha < 2$ and $q \approx 1/2$ for $\alpha > 2$.
Accordingly, in the last stage, $\alpha < 2$ corresponds to acceleration and $\alpha > 2$ corresponds to deceleration.
Consequently, when $\alpha <2$, the transition from deceleration to acceleration should be satisfied in the last stage. 
The transition requires $dq /d\tilde{a} <0$ and therefore, we examine the sign of $dq /d\tilde{a}$.
Differentiating $q$ given by Eq.\ (\ref{eq:q_power}) with respect to $\tilde{a}$ yields 
\begin{align}
\frac{dq}{d \tilde{a}} 
                 &=      \frac{  9   \Psi_{\alpha}  (1- \Psi_{\alpha})  \tilde{a}^{ - \frac{3(2-\alpha)}{2} -1  }    } {   4 \left [ (1- \Psi_{\alpha})  \tilde{a}^{ - \frac{3(2-\alpha)}{2}  }  + \Psi_{\alpha}    \right ]^{2}  }    ( \alpha -2 )   .
\label{eq:dqda}
\end{align}
This equation indicates that the sign of $dq /d\tilde{a}$ depends on $\alpha-2$ because $0 < \Psi_{\alpha} < 1$ is assumed.
From Eq.\ (\ref{eq:dqda}), we can confirm that $\alpha <2$ satisfies $dq /d\tilde{a} < 0$.

\section{Conclusions}
\label{Conclusions}

We studied irreversible entropy due to adiabatic particle creation in a flat FRW universe at late times.
To systematically examine such a dissipative universe, we phenomenologically formulated a dissipative cosmological model that includes a power-law term proportional to $H^{\alpha}$.
The irreversible entropy $S_{m}$ for the dissipative model was derived from an entropy relation for adiabatic particle creation.
In the dissipative universe, $S_{m}$ in the Hubble volume was found to be proportional to $H^{-1}$.
(The Bekenstein--Hawking entropy $S_{\rm{BH}}$ on the Hubble horizon is proportional to $H^{-2}$ in a flat FRW universe.)

Using the dissipative model, we examined the evolution of $S_{m}$ and the Bekenstein--Hawking entropy $S_{\rm{BH}}$, extending a previous analysis of a non-dissipative universe \cite{Koma14}.
The present dissipative model always satisfies the second law of thermodynamics for both $S_{m}$ and $S_{\rm{BH}}$, i.e., $\dot{S}_{m} \geq 0$ and $\dot{S}_{\rm{BH}} \geq 0$.
That is, the generalized second law of thermodynamics, i.e., $\dot{S}_{m} + \dot{S}_{\rm{BH}} \geq 0$, is also satisfied in the dissipative universe.

In addition, we examined the maximization of entropy, using the $(\tilde{a},\alpha)$ and $(\Psi_{\alpha},\alpha)$ planes. 
When $\alpha < 2$, the maximization of entropy for $S_{m}$, i.e., $\ddot{S}_{m} <0$, is always satisfied.
In contrast, even when $\alpha < 2$, $\ddot{S}_{\rm{BH}}<0$ is not satisfied in the early stage and should be satisfied in the last stage.
Therefore, constraints on $\ddot{S}_{\rm{BH}} <0$ are tighter than those on $\ddot{S}_{m} <0$.
Consequently, the maximization of total entropy depends almost entirely on the constraints on $\ddot{S}_{\rm{BH}} <0$.
The present study implies that the entropy maximization constrains the dissipative universe as if the universe behaves as an ordinary, isolated macroscopic system.
Note that Sol\`{a} and Yu have reported a similar result in a dissipative running-vacuum universe, in which $\alpha$ is set to be a constant value \cite{Sola2020}.

Furthermore, we examined constraints on an initially decelerating and then accelerating universe, which is implied by observations. 
When $\alpha <2$, a transition from deceleration to acceleration should be satisfied at least in the last stage. 
This constraint is likely consistent with the thermodynamic constraint. 
Cosmological observations should further constrain the dissipative universe, and these are left for future research.

\begin{acknowledgements}
The present study was supported by JSPS KAKENHI Grant Number JP18K03613.
The author wishes to thank the anonymous referee for very valuable comments which helped to improve this paper.
\end{acknowledgements}

\appendix

\section{Matter-dominated era} 
\label{MDE}

In Sec.\ \ref{Entropy Sm}, we derived the entropy density relation and the entropy relation, $s \propto H^{2}$ [Eq.\ (\ref{eq:ss0_HH0})] and $S_{m} \propto H^{-1}$ [Eq.\ (\ref{eq:SmSm0-HV})], for the present dissipative model.
In this appendix, the two relations are derived assuming a matter-dominated era (MDE) and $s \propto a^{-3}$.
In addition, the evolution of $S_{m}$ in the MDE model is examined.
Note that inflation of the early universe and the influence of radiation are not considered in this study.

From Eq.\ (\ref{eq:Sol_HH0_power}), the solution for the present dissipative model is written as
\begin{equation}  
    \left ( \frac{H}{H_{0}} \right )^{2-\alpha}  =   (1- \Psi_{\alpha})   \tilde{a}^{ - \frac{3(2-\alpha)}{2}  }  + \Psi_{\alpha}   .
\label{eq:Sol_HH0_power_2}
\end{equation}
The first term on the right-hand side of this equation is dominant both when $\alpha < 2$ and $\tilde{a} \ll 1$.
Accordingly, in the MDE (i.e., $\tilde{a} \ll 1$), this equation can be simplified and written as 
\begin{equation}  
             H^{2}  \propto    a^{-3}   \quad (\alpha <2) \quad \textrm{[MDE]}.
\label{eq:Sol_H_MDE}
\end{equation}
From Eq.\ (\ref{eq:NS-a3}), the entropy $S$ in the comoving volume is given by
\begin{equation}
           S = s a^3   .
\label{eq:NS-a3_2}
\end{equation}
The entropy density $s$ is assumed to be 
\begin{equation}
         s \propto a^{-3}  .
\label{eq:s-a-3_MDE}
\end{equation}
Using Eqs.\ (\ref{eq:Sol_H_MDE}) and (\ref{eq:s-a-3_MDE}), the entropy density relation is given by
\begin{equation}
         s \propto H^{2}   .
\label{eq:s-H2_MDE}
\end{equation}
Multiplying Eq.\ (\ref{eq:s-H2_MDE}) by the Hubble volume $V$ yields
\begin{equation}
         S_{m} \propto H^{-1}  ,
\label{eq:Sm-H_MDE}
\end{equation}
where $S_{m} = s V$ and $V \propto r_{H}^{3} \propto H^{-3}$ are used.
Equations (\ref{eq:s-H2_MDE}) and (\ref{eq:Sm-H_MDE}) are equivalent to Eqs.\ (\ref{eq:ss0_HH0}) and (\ref{eq:SmSm0-HV}), respectively.

In this way, we can obtain $s \propto H^{2}$ and $S_{m} \propto H^{-1}$, assuming the MDE and $s \propto a^{-3}$.
Of course, these assumptions should be invalid at the present time even if they are valid for the past.
That is, the evolution of the universe in the MDE gradually departs from that in the present dissipative model, with increasing $\tilde{a}$.
To examine this, we consider the  following MDE model.
Applying $\alpha <2$ and $\tilde{a} \ll 1$ to Eq.\ (\ref{eq:Sol_HH0_power_2}) yields
\begin{equation}  
    \left ( \frac{H}{H_{0}} \right )^{2}  =   (1- \Psi_{\alpha})^{\frac{2}{2-\alpha}}  \tilde{a}^{ - 3}     \quad (\alpha <2)  \quad \textrm{[MDE]}.
\label{eq:Sol_HH0_MDE}
\end{equation}
Substituting this equation into $S_{m}/S_{m,0} =(H/H_{0})^{-1}$ yields
\begin{equation}  
     \frac{S_{m}}{S_{m,0}}  =   (1- \Psi_{\alpha})^{\frac{1}{\alpha-2}}  \tilde{a}^{\frac{3}{2}}     \quad (\alpha <2)  \quad \textrm{[MDE]}.
\label{eq:Sm_MDE}
\end{equation}
When $\tilde{a}=1$, the two equations for the MDE model do not reduce to $1$, unlike for the present dissipative model.

\begin{figure} [t] 
\begin{minipage}{0.495\textwidth}
\begin{center}
\scalebox{0.3}{\includegraphics{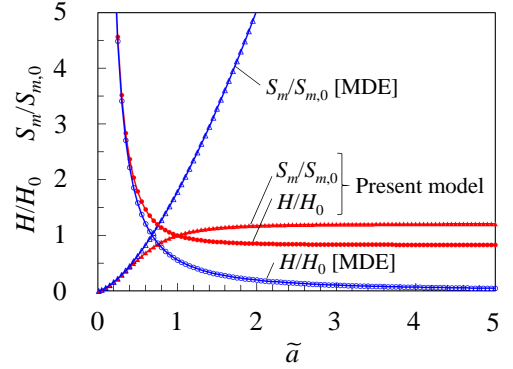}}
\end{center}
\end{minipage}
\caption{ (Color online). Evolution of $H/H_{0}$ and $S_{m}/S_{m,0}$ in the MDE and present dissipative models for $\Psi_{\alpha} =0.685$ and $\alpha =0$. 
The plots for the present model are from Figs.\ \ref{Fig-H-a}(a) and \ref{Fig-Sm_SBH}(a).
}
\label{Fig-H-Sm-a_MDE}
\end{figure} 

We now examine the evolution of the universe in the MDE and present dissipative models.
To this end, we set $\Psi_{\alpha} =0.685$ and $\alpha =0$.
Therefore, the evolution of the universe in the present dissipative model is equivalent to that in the $\Lambda$CDM model.
Figure\ \ref{Fig-H-Sm-a_MDE} shows the evolution of $H/H_{0}$ and $S_{m}/S_{m,0}$ in the two models.
When $\tilde{a} \ll 1$, $H/H_{0}$ for the two models agree with each other.
Similarly, $S_{m}/S_{m,0}$ for the two models agrees.
However, with increasing $\tilde{a}$, $H/H_{0}$ and $S_{m}/S_{m,0}$ for the MDE model depart from those for the present model.
In particular, the normalized $S_{m}$ for the MDE model rapidly increases with $\tilde{a}$, i.e., $S_{m} \propto \tilde{a}^{\frac{3}{2}}$, as shown in Fig.\ \ref{Fig-H-Sm-a_MDE} and Eq.\ (\ref{eq:Sm_MDE}).
Consequently, the MDE model does not satisfy the maximization of entropy.
In contrast, the normalized $S_{m}$ for the present model gradually approaches a constant value although this value increases with $\tilde{a}$.
The present dissipative model for $\alpha =0$ satisfies the maximization of entropy, as described in Sec.\ \ref{Entropy evolution}.

The above result implies that the two assumptions used here, i.e., that the MDE and $s \propto a^{-3}$, are valid at least when $\tilde{a} \ll 1$.
The MDE model should be useful for estimating the fundamental properties of cosmological models in a matter-dominated era.

\section{Bekenstein--Hawking entropy $S_{\rm{BH}}$ on the Hubble horizon for the present dissipative model} 
\label{Entropy on the horizon}

In this appendix, we examine the Bekenstein--Hawking entropy $S_{\rm{BH}}$ for the present dissipative model that includes $H^{\alpha}$ terms.
$S_{\rm{BH}}$ depends on the background evolution of the universe,
and the background evolution in the present model is equivalent to that in a $\Lambda (t)$ model with the $H^{\alpha}$ term, which was examined in a previous work \cite{Koma14}.
Therefore, we can use $S_{\rm{BH}}$ examined in Ref.\ \cite{Koma14}.
An expanding universe is assumed, as for the previous work.
For simplicity, equations for $\alpha \neq 2$ are shown here;
when $\alpha \rightarrow 2$, they reduce to those for $\alpha = 2$ \cite{Koma14}.

Using the result of Ref.\ \cite{Koma14}, 
$S_{\rm{BH}}$ for the present dissipative model is written as
\begin{align}  
S_{\rm{BH}}   &=  \frac{K}{H_{0}^{2}}     \left (     (1- \Psi_{\alpha})   \tilde{a}^{ - \frac{3(2-\alpha)}{2}  }  + \Psi_{\alpha}     \right )^{\frac{2}{\alpha-2}}      .
\label{eq:SBH_power_Complicate}      
\end{align}  
In the present paper, we use a normalized formulation.
The normalized $S_{\rm{BH}}$ is summarized as  
\begin{equation}  
   \frac{ S_{\rm{BH}}  }{S_{\rm{BH},0}  } =      
                                                              \left (     (1- \Psi_{\alpha})   \tilde{a}^{ - \frac{3(2-\alpha)}{2}  }  + \Psi_{\alpha}     \right )^{\frac{2}{\alpha-2}}     ,
\label{eq:SBHSBH0_power}
\end{equation}
where $S_{\rm{BH},0}$ is $S_{\rm{BH}}$ at the present time, which is given by $K/H_{0}^{2}$ from Eq.\ (\ref{eq:SBH2}).

Similarly, from Ref.\ \cite{Koma14}, we obtain the first derivative of $S_{\rm{BH}}$, i.e., $\dot{S}_{\rm{BH}} $, which is written as  
\begin{align}  
\dot{S}_{\rm{BH}}   
&=  \frac{3K}{H_{0}}  \left (  1 -   \frac{ \Psi_{\alpha} }{ (1- \Psi_{\alpha})   \tilde{a}^{ - \frac{3(2-\alpha)}{2}  }  + \Psi_{\alpha}    }  \right )       \notag \\
& \quad \times  \left [ (1- \Psi_{\alpha})   \tilde{a}^{ - \frac{3(2-\alpha)}{2}  }  + \Psi_{\alpha}  \right ]^{ \frac{1}{\alpha-2} }  .
\label{eq:dSBH_2_3_power_Complicate}      
\end{align}  
This equation can be written as
\begin{align}  
\dot{S}_{\rm{BH}}   
&=  \frac{3K}{H_{0}}   \frac{ (1- \Psi_{\alpha})   \tilde{a}^{ - \frac{3(2-\alpha)}{2}  }   }{  \left [ (1- \Psi_{\alpha})   \tilde{a}^{ - \frac{3(2-\alpha)}{2}  }  + \Psi_{\alpha}  \right ]^{ \frac{3- \alpha}{2-\alpha} }  }        .
\label{eq:dSBH_2_3_power_Complicate2}      
\end{align}  
Using $S_{\rm{BH},0}  = K/H_{0}^{2}$, the normalized $\dot{S}_{\rm{BH}} $ is written as 
\begin{equation}  
   \frac{ \dot{S}_{\rm{BH}}  }{S_{\rm{BH},0} H_{0} } =     
                                                              \frac{ 3 (1- \Psi_{\alpha})   \tilde{a}^{ - \frac{3(2-\alpha)}{2}  }   }{  \left [ (1- \Psi_{\alpha})   \tilde{a}^{ - \frac{3(2-\alpha)}{2}  }  + \Psi_{\alpha}  \right ]^{ \frac{3- \alpha}{2-\alpha} }  }     .
\label{eq:dSBHSBH0_power}
\end{equation}

In addition, we obtain the second derivative of $S_{\rm{BH}}$ from Ref.\ \cite{Koma14}.
$\ddot{S}_{\rm{BH}}$ for the present dissipative model is written as 
\begin{align}  
 \ddot{S}_{\rm{BH}}   &= \frac{9K}{2}   \left (  1 -  \frac{  \Psi_{\alpha}                      }{   (1- \Psi_{\alpha})    \tilde{a}^{ - \frac{3(2-\alpha)}{2}  }  + \Psi_{\alpha}         }  \right )    \notag \\
                               & \quad \times           \left [  1 -   \frac{  \Psi_{\alpha}  (3 - \alpha )  }{   (1- \Psi_{\alpha})   \tilde{a}^{ - \frac{3(2-\alpha)}{2}  }  + \Psi_{\alpha}         }  \right ]   .
\label{eq:d2Sdt2_power_1_Complicate}
\end{align}
This equation can be written as
\begin{align}  
 \ddot{S}_{\rm{BH}}   &= \frac{9K}{2}    \frac{  (1- \Psi_{\alpha})    \tilde{a}^{ - \frac{3(2-\alpha)}{2}  }                     }{   (1- \Psi_{\alpha})    \tilde{a}^{ - \frac{3(2-\alpha)}{2}  }  + \Psi_{\alpha}         }      \notag \\
                               & \quad \times       \frac{  (1- \Psi_{\alpha})    \tilde{a}^{ - \frac{3(2-\alpha)}{2} }    + (\alpha-2)\Psi_{\alpha}            }{   (1- \Psi_{\alpha})    \tilde{a}^{ - \frac{3(2-\alpha)}{2}  }  + \Psi_{\alpha}         }         \notag \\
                            &= \frac{9K}{2}    \frac{  (1- \Psi_{\alpha})    \tilde{a}^{ - \beta  }       \left [  (1- \Psi_{\alpha})    \tilde{a}^{ - \beta }    + (\alpha-2)\Psi_{\alpha}   \right ]               }{   \left [    (1- \Psi_{\alpha})   \tilde{a}^{ - \beta }  + \Psi_{\alpha}         \right ]^{2}      }      ,
\label{eq:d2Sdt2_power_1_Complicate2}
\end{align}
where $\beta$ is given by
\begin{equation}  
                                 \beta  = \frac{3(2-\alpha)}{2} .
\label{eq:beta}
\end{equation}
The normalized $\ddot{S}_{\rm{BH}}$ is written as
\begin{equation}  
   \frac{ \ddot{S}_{\rm{BH}}  }{S_{\rm{BH},0} H_{0}^{2} } =      
                                                               \frac{9}{2}    \frac{  (1- \Psi_{\alpha})    \tilde{a}^{ - \beta }     \left [  (1- \Psi_{\alpha})    \tilde{a}^{ - \beta }    + (\alpha-2)\Psi_{\alpha}   \right ]             }{   \left [    (1- \Psi_{\alpha})   \tilde{a}^{ - \beta  }  + \Psi_{\alpha}         \right ]^{2}      }      .
\label{eq:d2SBH2SBH0_power}
\end{equation}
For details of the derivations, see Ref.\ \cite{Koma14}.

\end{document}